\pdfoutput=1
\documentclass[aps,pre,twocolumn, amsfonts,showpacs,superscriptaddress]{revtex4-1}

\usepackage{mathptmx}
\usepackage{epsfig,amsopn}
\usepackage{graphicx}
\usepackage{amsmath,amssymb}
\usepackage{natbib}
\usepackage{braket}
\usepackage{bm}

\newcommand{\eref}[1]{Eq.~(\ref{#1})}%
\newcommand{\fref}[1]{Fig.~\ref{#1}} %
\newcommand{\Fref}[1]{Figure~\ref{#1}}%
\newcommand{\Sref}[1]{Section~\ref{#1}}%

\newcommand{\sgn}[1]{\mathrm{sgn}({#1})}%
\newcommand{\erfc}{\mathrm{erfc}}%

\begin{document}

\title{Random walk with random resetting to the maximum}

\author{Satya N. Majumdar}
\address{LPTMS, CNRS, Univ Paris Sud, Universit\'e Paris-Saclay, 91405 Orsay, France}

\author{Sanjib Sabhapandit}
\address{Raman Research Institute, Bangalore 560080, India}

\author{Gr\'egory Schehr}
\address{LPTMS, CNRS, Univ Paris Sud, Universit\'e Paris-Saclay, 91405 Orsay, France}
\date{\today}

\begin{abstract} We study analytically a simple random walk model on a 
one-dimensional lattice, where at each time step the walker resets to the 
maximum of the already visited positions (to the rightmost visited site) 
with a probability $r$, and with probability $(1-r)$, it undergoes 
symmetric random walk, i.e., it hops to one of its neighboring 
sites, with equal probability $(1-r)/2$. For $r=0$, it reduces to a 
standard random walk whose typical distance grows as $\sqrt{n}$ for large 
$n$. In presence of a nonzero resetting rate $0<r\le 1$, we find that both 
the average maximum and the average position grow ballistically for large 
$n$, with a common speed $v(r)$. Moreover, the fluctuations around their 
respective averages grow diffusively, again with the same diffusion 
coefficient $D(r)$. We compute $v(r)$ and $D(r)$ explicitly. We also show 
that the probability distribution of the difference between the maximum 
and the location of the walker, becomes stationary as $n\to \infty$.  
However, the approach to this stationary distribution is accompanied by a 
dynamical 
phase transition, characterized by a weakly singular large 
deviation 
function. We also show that $r=0$ is a special `critical' point, 
for which 
the growth laws are different from the $r\to 0$ case and we calculate the 
exact crossover functions that interpolate between the critical $(r=0)$ 
and the off-critical $(r\to 0)$ behavior for finite but large $n$.

\end{abstract}
\pacs{05.40.-a, 02.50.-r,87.23.Ge}
\maketitle

\section{Introduction}

Search problems appear in diverse contexts~\cite{Viswanathan:1996fr,
Viswanathan:1999kf, Edwards:2007gq, Benichou:2011fy} and there have
been a recent surge of interests in the physics community in these
problems~\cite{Luz:2009bb}. Search strategies may be either
systematic or random. In the systematic strategies, the searcher uses
deterministic rules (e.g., `lawnmower') to find a target. On the other
hand, the random search mechanism typically involves two kinds of
moves: local steps when the searcher looks for a target, and
long-range moves during which the searcher does not look for the
target but relocates itself to a different territory. The slow search
phase is typically modelled by a diffusion or a random walk. The
long-range moves may be modelled depending on the specific 
application~\cite{Benichou:2011fy}. 

A particularly simple long-range strategy consists of `resetting' the
searcher to a fixed location (say to the initial starting point)
with a finite probability/rate. The rational behind this strategy
is that if one does not succeed in finding the target via short-range
diffusion, it is better to `restart' the process, rather than continuing
on the short-range moves. 
The effect of such stochastic resetting
was first studied by Manrubia and Zanette~\cite{Manrubia:1999} in  
the context of multiplicative processes and a slightly different version was studied later by 
Gelenbe~\cite{Gelenbe:2010} in the context of network theory.
Such `restart' strategy also plays an important role in 
randomized search algorithms for combinatorial optimization 
problems~\cite{Montanari:2002,Janson:2012}. 
 
Recently, a very simple
model of a Brownian searcher in presence of stochastic resetting
to its initial position with rate $r$ was     
introduced by Evans and Majumdar~\cite{Evans:2011jo}.
In presence of a nonzero $r$, it was shown that at long times, the probability
distribution of the position of the walker reaches a non-equilibrium
steady state~\cite{Evans:2011jo}. The temporal relaxation to this
steady state was also studied recently~\cite{Majumdar:2015} and
an interesting dynamical phase transition was found: as time progresses,
an inner core region around
the resetting point reaches the steady state, while the region outside the 
core is still transient. The boundaries of
the core region grow linearly with time at late times~\cite{Majumdar:2015}.

In presence of resetting with rate $r$, the mean first-passage time to 
find a target located at the origin, by a searcher starting and resetting to 
$x_0$, was computed exactly~\cite{Evans:2011jo} and was found to have
a minimum at an optimal resetting rate $r^*$, thus making the search
process efficient in presence of resetting. This conclusion holds
in all dimensions~\cite{Evans:2014hd}. Also, it was proved that
this non-equilibrium reset dynamics is more efficient in target search
compared to an equilibrium Langevin dynamics in presence of an external
potential leading to the same steady state~\cite{Evans:2013fd}.  

This simple model of diffusion with stochastic resetting, in
the single searcher setting, has been generalized in various ways. For example,
when the target as well as the resetting positions are not fixed
but drawn from specified probability distributions~\cite{Evans:2011el},
in presence of partial detection (or absorption) of the target by a 
searcher~\cite{Whitehouse:2013ds},
when the searcher performs a continuous-time random 
walk~\cite{Montero:2013} or a L\'evy flight instead of a 
Brownian motion/random walk~\cite{Kusmierz:2014,Kusmierz:2015}, when the searcher
moves in a bounded domain~\cite{Christou:2015} or in the
presence of a confining potential~\cite{Pal:2015}, when the resetting
occurs to any of the previously visited sites with a rate proportional to the
number of visits to the site~\cite{Boyer:2014} etc.
Recently the model of random walks with resetting has also
been used to understand the behavior in models of
enzymatic reactions in biology~\cite{Reuveni:2014}. 

Going beyond the one particle setting,
the effect of the resetting mechanism in searching an immobile
target has also been studied in presence of multiple, but non-interacting
searchers~\cite{Evans:2011jo}. More recently, the resetting has been 
studied in
spatially extended many-body interacting systems, such as for fluctuating 
interfaces~\cite{Majumdar:2015,Gupta:2014}
as well as a class of reaction-diffusion 
models~\cite{Durang:2014}---in both cases,
the natural dynamics of the system is stochastically interrupted
by resetting it to the initial configuration at a nonzero rate $r$.
A nonzero $r$ leads to new non-equilibrium stationary states in such
extended systems~\cite{Majumdar:2015,Gupta:2014}.
    
In this paper, we consider a model where the searcher remembers the
maximum location visited so far and the long-range move consists in
resetting to this current maximum. This strategy may be thought as the mixture
of the systematic search and the random search. In the systematic
search strategy since each location is visited only once, if a target
in an already visited place is missed (by an \textit{imperfect
searcher}), it is never going to be detected. In the new strategy
discussed in this paper, the searcher revisits already searched
locations (with certain probability), but also feels
a dynamical bias towards exploring new locations (by resetting to the maximum).

This model may actually be useful in the context of animals searching for
food. During the foraging period, it is well known that an animal typically 
performs a random walk in search of food~\cite{Berg-book,Bart:2005}. 
It is however quite natural 
for an intelligent animal (with memory) to remember the already visited 
(explored) sites and thus 
to have a natural tendency to relocate once in a while 
to the frontier between already explored and yet unexplored territories, where
the probability of finding food may be higher due to the
proximity of the unexplored territory. 
In a one dimensional setting, the animal, besides short-range diffusion,
may relocate with a nonzero probability to the current maximum or 
to the current minimum, which together
constitute the frontier between explored and unexplored territories.
In this paper, we consider an 
even more simplified {\em {directed}} version that has the advantage of being
exactly solvable. In 
our model the
animal starting at the origin, besides performing short-range 
standard random walk,
relocates stochastically with a nonzero probability to
only the positive side of the frontier, i.e., the farthest
visited site so far to the right of the origin (i.e., to the 
maximum). We will see
that despite the fact that the position of the walker evolves
via a non-Markovian dynamics (as it remembers the maximum position
so far in order to relocate), the model allows for an exact solution
and thus provides interesting insights into this long-range search strategy.

The rest of the paper is organized as follows. In 
\Sref{summary} we introduce the model precisely and  
summarize our main exact results.
In \Sref{generating function} we derive the
generating functions for the probabilities of the position and the
maximum after $n$ time steps. We then examine in detail the
asymptotic large $n$ statistics
in the 
two opposite limits 
respectively in the next two sections:
(i) the case without resetting but only with diffusion in \Sref{r=0
case} and (ii) the case where only resetting occurs without any
diffusion in \Sref{r->1 limit}.
In \Sref{general resetting} we analyze the situation 
for arbitrary resetting rate.
In \Sref{scaling-limit}, we analyze the statistics
of the maximum and the position in the scaling limit $r\to 0$,
$n\to \infty$ while keeping the product $r\, n$ fixed.
The crossover scaling functions are computed exactly in this Section
and compared to numerical simulation results.
Finally we conclude with a summary and some open questions in 
\Sref{conclusion}.

\section{The model and the summary of main results}
\label{summary}

We consider a walker 
moving on a one-dimensional lattice, initially
starting from the origin. Here each lattice site should be thought of
as a `region' which is much larger than the `size of the searcher',
but much smaller than the whole region. The searcher spends some
characteristic time $\tau$ in each region (lattice site), and we
consider time steps in units of $\tau$ and take it to be discrete.
Let $x(n)$ denote the position of the walker at step $n$ and
$m(n)$ denote the current position of the maximum at step $n$, i.e.
\begin{equation}
m(n)= {\rm max}\left[x(0)=0,\, x(1),\, x(2),\ldots, x(n)\right].
\label{max_def}
\end{equation}

\begin{figure*}
\begin{center}
\includegraphics[width=.9\hsize]{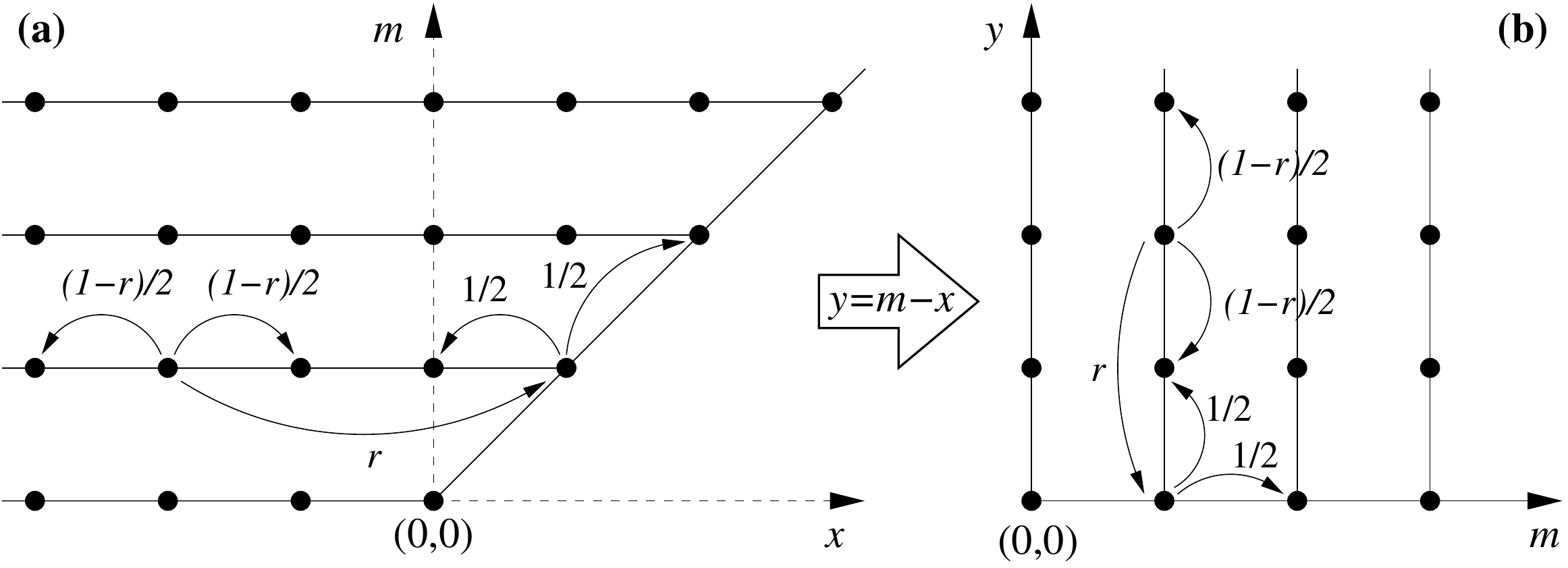}
\caption{{\bf (a)} The allowed lattice points for the walker to move
  in the $(x,m)$ plane. The walker is restricted to move only along
  the horizontal (constant $m$) lines except when it is on the $x=m$
  line. When on the $x=m$ line, the walker can either move to a level
  up on the same line $(m,m)\to (m+1,m+1)$ with probability $1/2$ or
  move to the left site while staying on the same level $(m,m)\to
  (m-1,m)$ with probability $1/2$. {\bf (b)} The comb lattice
  structure after the transformation $y=m-x$.  In the $(y,m)$ plane,
  the walker is restricted to move along the vertical lines except
  when $y=0$. From a point $(0,m)$ the walker can go either to
  $(0,m+1)$ with probability $1/2$ or to $(1,m)$ with probability
  $1/2$.  }
\label{lattice}
\end{center}
\end{figure*}

The position $x(n)$ evolves with time via the following stochastic
dynamics.  At any given time step $n$, if the position $x(n)$ of the
walker is less than the maximum position $m(n)$ reached up to that
time (i.e, $x(n)<m(n)$ strictly), then in the next time step, the
position is reset to the maximum position with probability $r$. With
the remaining probability $(1-r)$, the walker moves either to the
right or to the left lattice site, with equal probability $(1-r)/2$.
On the other hand, if $x(n)=m(n)$, then in the next time step, the
walker moves either to the right or to the left lattice site with
equal probability $1/2$. The dynamics is precisely defined by the
following evolution rules~[see \fref{lattice}~(a)]: \\

\noindent if $x(n)<m(n)$
\begin{equation}
(x,m) \to
\begin{cases}
 (x+1,m) &\text{with probability $(1-r)/2$}, \\
 (x-1,m) &\text{with probability $(1-r)/2$},\\
 (m,m) &\text{with probability $r$},
\end{cases}
\label{evolution.1}
\end{equation}
and if $x(n)=m(n)$,  
\begin{equation}
(x,m) \to
\begin{cases}
 (m+1,m+1) & \text{with probability $1/2$}, \\
 (m-1,m) & \text{with probability $1/2$}.
\end{cases}
\label{evolution.2}
\end{equation}
Evidently, the evolution of $x(n)$ is non-Markovian by itself, 
since
the walker has to remember the maximum position reached so far in order
to reset. However, the dynamics of the pair of stochastic variables 
$\{x(n),m(n)\}$ is Markovian in the two dimensional $(x,m)$ plane, and
this is the key point behind the solvability of the model. 
\Fref{lattice} depicts
the motion in the $(x,m)$ plane.

It is useful to summarize our main results. Our main objective is
to compute the statistics of the two random variables $x(n)$ and $m(n)$.
Let us first recall that in the absence of resetting ($r=0$), the walker 
performs a standard one 
dimensional random walk, for which $x(n)$, converges
to a Gaussian random variable with zero mean and variance $n$, for large $n$.
Hence, the probability distribution of the position converges, for large $n$,
to $P_{\rm x}(x,n) \to \sqrt{\frac{1}{2\pi n}}\, \exp\left[-x^2/{2n}\right]$.
Similarly, for $r=0$, the distribution of the maximum $m(n)\ge 0$, 
for large $n$, 
converges to a half-Gaussian: 
$P_{\rm m}(m,n)\to \sqrt{\frac{2}{\pi n}}\,\exp\left[-m^2/2n\right]$ with
support only over $m\ge 0$.
Thus, for $r=0$, both the position and the maximum typically grow diffusively 
as $\sqrt{n}$ for large $n$.  

When the resetting to the maximum is switched on ($r>0$), the walker
feels a dynamical bias towards the maximum. Hence, one expects that
both $x(n)$ and $m(n)$ will grow faster than pure diffusion for large $n$.
The question is how much faster? We will see that for $r>0$ (strictly),
both $m(n)$ and $x(n)$ grow linearly with $n$ for large $n$ with
the {\it same} speed. In addition, the variance of both $m(n)$ and $x(n)$
grow diffusively for large $n$ with the {\it same} diffusion coefficient.
This suggests that for all $r>0$, the position latches on to the maximum
and indeed, we show that the difference variable $m(n)-x(n)$ approaches
a stationary distribution as $n\to \infty$ for all $r>0$.
Our exact results are summarized below. \\

\noindent {\bf Statistics of the maximum $m(n)$:} The average maximum, for 
large $n$, behaves as
\begin{align}
\label{r0meanm}
\langle m(n)\rangle &\simeq  \sqrt{\frac{2\,n}{\pi}} &{\rm for}\quad
r=0, 
 \\
\label{rpmeanm}
& \simeq  v(r)\, n  &{\rm for} \quad r>0, 
\end{align}
where the speed $v(r)$ is given by 
\begin{equation}
v(r)=\frac{r (1-r) }{r-2 r^2+\sqrt{r (2-r)}}.
\label{v(r)}
\end{equation}
The speed vanishes as $v(r)\approx \sqrt{r/2}$ as $r\to 0$. 
The variance of $m(n)$ grows diffusively for large $n$,
\begin{equation}
\sigma_m^2= \langle m^2(n)\rangle -{\langle m(n)\rangle}^2\simeq
  D_m(r)\, n, 
\end{equation}
where the diffusion coefficient (the subscript $m$ in $D_m(r)$ denotes
the random variable $m$) is given by
\begin{align}
 \label{r0mdiff}
D_m(r) & = \left(1-\frac{2}{\pi}\right) &{\rm for}
\quad r=0,  \\
\label{rpmdiff}
& =  
D(r)  &{\rm for}\quad r>0, 
\end{align}
where $D(r)$, for $r>0$, is given by
\begin{eqnarray}
&&D(r)=\frac{(1-r) r^2 }{\sqrt{r (2-r)} \left[r-2
    r^2+\sqrt{r(2-r)}\right]^3} \label{D(r)} \\
&&\times \left[(2-2 r-5 r^2+3 r^3)+ (2-r-r^2+2
  r^3) \sqrt{r(2-r)} \right] \;. \nonumber
\end{eqnarray}

\begin{figure}
\centerline{\includegraphics[width=.95\hsize]{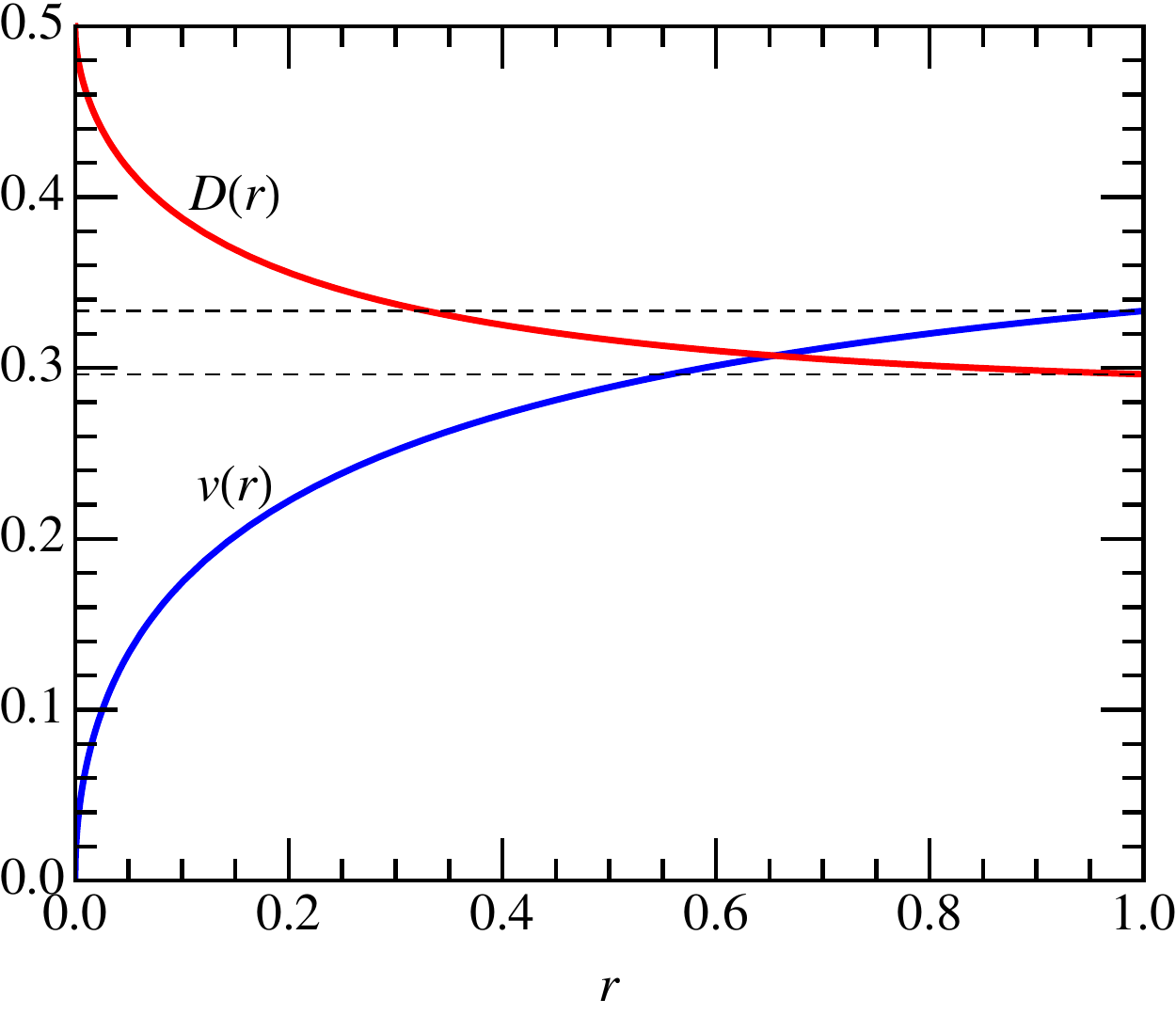}}
\caption{\label{v&D} (Color online) $v(r)$ and $D(r)$ as a function of
  $r$ (in blue and red respectively). The
  dashed lines show their limiting values, $1/3$ and $8/27$
  respectively, for $r\to 1$.}
\end{figure}

The behavior of $v(r)$ and $D(r)$, vs. $r$, are shown 
in \fref{v&D}.
Note that as $r\to 0$, $D(r)\to 1/2\neq D_m(0)=(1-2/\pi)$. Thus, there is
a discontinuity in the variance as $r\to 0$. 
These results clearly indicate that $r=0$ is a {\em {singular/critical}}
point. Indeed, we find that near the critical point $r=0$, there is
a scaling regime. Taking $r\to 0$, $n\to \infty$, but keeping the
product $rn$ fixed, we find that the mean 
and the variance exhibit
the following scaling behavior
\begin{align}
\label{meanmscaling}
\langle m(n)\rangle &\to  \sqrt{n}\, f_m(rn) , 
\\
\label{varmscaling}
\sigma_m^2 &\to  n\, F_m(rn) , 
\end{align}
where the two scaling functions $f_m(y)$ and $F_m(y)$
have nontrivial expressions
\begin{align}
\label{fm}
f_m(y) &= \frac{1}{\sqrt{2y}}\left[\left(y+\frac{1}{2}\right)\, {\rm erf}
\left(\sqrt{y}\right)+ \sqrt{\frac{y}{\pi}}\, e^{-y}\right]\, , \\
\label{Fm}
F_m(y)&= 1+\frac{y}{2}-f_m^2(y)\, ,  
\end{align}
where ${\rm erf}(z)= \frac{2}{\sqrt{\pi}}\, \int_0^z e^{-u^2}\, du$,
is the error function.  The scaling function $f_m(y)$ have the
asymptotic behaviors: $f_m(y)\sim \sqrt{2/\pi} + \sqrt{2/\pi}(y/3)$ as
$y\to 0$, and $f_m(y) \sim \sqrt{y/2}+1/\sqrt{8y}$ as $y\to
\infty$. Consequently, $F_m(y)\to (1-2/\pi)$ as $y\to 0$, and
$F_m(y)\to 1/2$ as $y\to \infty$.  These two limiting behaviors of the
scaling functions (for the mean and the variance) then smoothly
interpolate between $r=0$ (strictly) and $r>0$ (and $n\to \infty$).
For any small but fixed $r$, there is a crossover time $n^*(r)\sim
1/r$, such that for $n<n^*(r)$, the mean and the variance grow with
$n$ simply as a random walk ($r=0$): $\langle m(n)\rangle \sim
\sqrt{2n/\pi}$ and $\langle \sigma_m^2\rangle \sim
(1-2/\pi)\,n$. However, for $n>n^*(r)$, the walker starts sensing the
presence of a finite resetting rate $r$ and crosses over to a new
behavior where the mean maximum grows linearly with $n$, $\langle
m(n)\rangle \sim v(r)\, n$ and the variance grows diffusively,
$\sigma_m^2\sim D(r) \, n$ with the diffusion constant $D(r)$ given in
Eq. (\ref{D(r)}).  \\

\noindent {\bf Statistics of the position $x(n)$:} We find that the
mean position behaves as 
\begin{align}
 \label{r0meanx}
\langle x(n)\rangle & = 0  &\text{for}\quad r=0, \\
 \label{rpmeanx}
& \simeq  v(r)\, n  &\text{for}\quad r>0, 
\end{align}
where the speed $v(r)$, for $r>0$, is the same as that of the maximum
given in Eq. (\ref{v(r)}).
The variance of $x(n)$ grows diffusively for large $n$, 
\begin{equation}
\sigma_x^2 =  \langle x^2(n) \rangle- {\langle x(n)
\rangle}^2 \simeq D_x(r)\, n, 
\end{equation}
with the diffusion coefficient (the subscript $x$ denotes the random
variable $x$)
\begin{align}
 \label{r0diffcx}
D_x(r) &=   1  &{\rm for}\quad r=0, \\
\label{rpdiffcx}
& =  D(r)  &{\rm for} \quad r>0,
\end{align}
where the diffusion coefficient $D(r)$ is the same as that of the maximum
and is given in Eq. (\ref{D(r)}). 
As in the case of the maximum, the diffusion
coefficient $D(r\to 0)=1/2\neq D_x(0)=1$ undergoes a discontinuous jump
at the critical point $r=0$. Similar to $m(n)$, the behavior
of the mean and the variance of $x(n)$ exhibit a scaling behavior in
the scaling regime $r\to 0$, $n\to \infty$ with the product $rn$ fixed 
\begin{align}
 \label{meanxscaling}
\langle x(n)\rangle &\to  \sqrt{n}\, f_x(rn), \\
\label{varxscaling}
\sigma_x^2 &\to  n\, F_x(rn)\,. 
\end{align}
The scaling functions have the exact expressions
\begin{align}
\label{fx}
f_x(y) & = 
\frac{1}{\sqrt{2y}}\left[\left(y-\frac{1}{2}\right)\, {\rm erf}
\left(\sqrt{y}\right)+ \sqrt{\frac{y}{\pi}}\, e^{-y}\right]\,, \\
\label{Fx}
F_x(y)&= \frac{y}{2}+\frac{1-e^{-y}}{y}-f_x^2(y)\,.  
\end{align}
The scaling function $f_x(y)$ has the asymptotic behavior: 
$f_x(y)\sim (2/3) \sqrt{2/\pi}\, y - (2/15)\sqrt{2/\pi}\,y^2$ as $y\to 0$ and $f_x(y)\sim \sqrt{y/2}-1/\sqrt{8y}$
as $y\to \infty$. As a result, $F_x(y)\to 1$ as $y\to 0$ and $F_x(y)\to 1/2$
as $y\to \infty$. These scaling functions then smoothly interpolate
between the critical ($r=0$) and off-critical ($r>0$) growth of the
mean and the variance. \\


\noindent{\bf Statistics of the difference $y(n)=m(n)-x(n)$:}
We show that the probability $Q_\mathrm{y}(y,n)$ that the position at the 
$n$-th
step is at a distance $y$ away from the global maximum, has the large
deviation form
\begin{equation}
Q_\mathrm{y}(y=wn,n)  \sim
\exp\bigl[-n H(w)\bigr],
\label{Q_y large}
\end{equation}
where the large deviation function is given by
\begin{equation}
H(w)=
\begin{cases}\displaystyle
w\ln\left[\frac{1+\sqrt{r(2-r)}}{1-r}\right] &\text{for~ $w<w^*$},\\[5mm]
\displaystyle
\frac{w}{2}\ln\frac{1+w}{1-w} + \ln
  \frac{\sqrt{1-w^2}}{1-r}   &\text{for~ $w> w^*$},
\end{cases}
\label{H(w)}
\end{equation}
with $w^*=\sqrt{r(2-r)}$. This result shows that for a given $n$, 
the probability $Q_{\rm y}(y,n)$ becomes independent of $n$ for
$y< w^* n$:
\begin{equation}
Q_{\rm y}(y,n)\sim \exp\left[-\ln\left(\frac{1+\sqrt{r(2-r)}}{1-r}\right)\, 
y\right],
\label{qyless}
\end{equation}
while for $y>w^* n$, the distribution $Q_{\rm y}(y,n)$ is still time-dependent.
In other words, the distribution of $y$ becomes stationary on a larger
and larger length scale $y^*(n)= w^* n$ that grows linearly with time $n$.
Moreover, the rate function $H(w)$ is weakly non-analytic at $w=w^*$:
the second derivative $H''(w)$ is discontinuous at $w=w^*$. This
signals a {\it dynamical} phase transition, similar to the one  
observed in the
temporal evolution of the distribution of position of a Brownian motion
with resetting to its initial position~\cite{Majumdar:2015}.

\section{The derivation using generating functions}
\label{generating function}

In this section we outline the derivation of our results. 
We start with the dynamics of the two basic observables
$x(n)$ and $m(n)$ given in Eqs. (\ref{evolution.1}) and
(\ref{evolution.2}).
Since, we have $m \ge 0$ and $x \le m$, it is convenient to define 
the difference variable $y=m-x$, where $y\ge 0$. In terms of $y$, the 
dynamics in Eqs. (\ref{evolution.1}) and (\ref{evolution.2}) get
translated into the equivalent forms [see \fref{lattice}~(b)]:\\

\noindent if $y> 0$
\begin{equation}
(y,m) \to
\begin{cases}
 (y-1,m) & \text{with probability $(1-r)/2$}, \\
 (y+1,m) & \text{with probability $(1-r)/2$},\\
 (0,m) & \text{with probability $r$},
\end{cases}
\label{yevol.1}
\end{equation}
and if $y=0$ 
\begin{equation}
(y,m) \to
\begin{cases}
 (0,m+1) & \text{with probability $1/2$}, \\
 (1,m) & \text{with probability $1/2$}.
\end{cases}
\label{yevol.2}
\end{equation}


Let $P(x,m,n)$ and $Q(y,m,n)$ denote the joint probability distribution of
$(x, m)$ and $(y, m)$ respectively, at the $n$-th time
step. Evidently, $P(x,m,n)=Q(m-x,m,n)$.
Using the dynamics in Eqs. (\ref{yevol.1}) and (\ref{yevol.2}), it is easy 
to write down the master equation for $Q(y,m,n)$ as
\begin{align}
Q(y,m,n) = \left[\frac{1-r}{2} +\frac{r}{2} \delta_{y,1} \right]
Q(y-1,m,n-1) \notag\\+ \frac{1-r}{2} Q(y+1,m,n-1),
\label{masteryp}
\end{align}
for $y>0$,  and 
\begin{align}
Q(0,m,n)=\frac{1-r}{2} Q(1,m,n-1) +\frac{1}{2} Q(0,m-1,n-1) \notag\\
+r \sum_{y=1}^\infty Q(y,m,n-1),
\label{mastery0}
\end{align}
with the initial condition $Q(y,m,0)=\delta_{y,0}\delta_{m,0}$, and
the boundary conditions $Q(y\to\infty,m,n)=0$ and
$Q(y,m\to\infty, n)=0$.

To solve the set of linear equations (\ref{masteryp}) and 
(\ref{mastery0}), it is natural to define the generating 
function
\begin{equation}
G(s,z,\lambda)=\sum_{y=0}^\infty \sum_{m=0}^\infty
\sum_{n=0}^\infty
Q(y,m,n) s^y z^m  \lambda^n.
\end{equation}
Evidently, $G(1,z,\lambda)$ is the generating function for the
probability distribution of the global maximum position and
$G(s,1,\lambda)$ is the generating function for the probability
distribution of the position. Moreover, $G(1,1,\lambda)$ must be equal
to $(1-\lambda)^{-1}$ as demanded by the normalization of
the probability.  
After straightforward algebra, it follows that $G(s,z,\lambda)$ satisfies 
\begin{align}
&G(s,z,\lambda) \left[1-\frac{a}{2} \left(s+\frac{1}{s}\right)\right]
=\notag\\
&1 + \left(rs-\frac{1-r}{s}+z-2r\right)\frac{\lambda}{2}
F(z,\lambda) + r \lambda G(1,z,\lambda),
\label{Gsz}
\end{align}
where 
\begin{equation}
\label{a}
a=(1-r) \lambda
\end{equation}
and 
\begin{equation}
F(z,\lambda)=\sum_{n=0}^\infty \sum_{m=0}^\infty Q(0,m,n) z^m \lambda^n.
\end{equation}
The expression for $G(1,z,\lambda)$ can be obtained by setting $s=1$
in \eref{Gsz} as
\begin{equation}
G(1,z,\lambda) = \frac{1}{1-\lambda} \left[1-(1-z)\frac{\lambda}{2}
  F(z,\lambda) \right].
\label{G1z}
\end{equation}
The normalization condition $G(1,1,\lambda)=(1-\lambda)^{-1}$ is
immediately checked from above. Substituting $G(1,z,\lambda)$ in
\eref{Gsz}, after some algebra, we get
\begin{align}
G(s,z,\lambda) &  \frac{a}{2 s} (s_+-s) (s-s_-) =\notag\\
&\frac{1-a}{1-\lambda}
\left( 1- \bigl[z_1(s) -z\bigr]\frac{\lambda}{2}F(z,\lambda) \right), 
\label{Gsz-2}
\end{align}
where
\begin{equation}
\label{s_pm}
s_\pm=\frac{1\pm\sqrt{1-a^2}}{a}
\end{equation}
and
\begin{equation}
\label{z1}
z_1(s)
=\frac{1-\lambda}{1-a}\left[-rs + \frac{1-r}{s} +2r
  +\frac{r\lambda}{1-\lambda} \right].
\end{equation}

In Eq. (\ref{Gsz-2}), the function $F(z,\lambda)$ is still undetermined
and has to be determined self-consistently. To proceed, we note 
from \eref{Gsz-2} that $G(s,z,\lambda)$ has two poles at
$s=s_\pm$ respectively. Therefore, inverting with respect to $s$,
gives the form
\begin{equation}
\sum_{\lambda=0}^\infty 
\sum_{m=0}^\infty
Q(y,m,n) z^m \lambda^n = \frac{A}{s_+^y} + \frac{B}{s_-^y}~, 
\label{poleAB}
\end{equation}
where $A$ and $B$ are the residues at the two poles.
However, from \eref{s_pm}, we notice that $s_+ > 1$ and $s_- < 1$. Hence, the
$(1/s_-)^{y}$ term in the above expression diverges when
$y\to\infty$, which is inconsistent with the boundary
condition $Q(y\to\infty,m,n)=0$. The only way to prevent
this blow up is that $B$ must 
necessarily vanish, 
which implies that the right hand side of \eref{Gsz-2}
vanishes for $s=s_-$. We note that this method of determining the 
self-consistency condition
via the `pole-cancelling' mechanism was used before in other 
contexts~\cite{Rajesh1:2000,Rajesh2:2000,Rajesh:2001}. 
This condition then determines $F(z,\lambda)$ as
\begin{equation}
\frac{\lambda}{2}F(z,\lambda)=\frac{1}{z_0-z},
\qquad
\mathrm{where}~~ z_0=z_1(s_-). 
\label{Fz}
\end{equation}
Substituting $F(z,\lambda)$ in \eref{G1z}, gives
the generating function of the maximum as
\begin{equation}
\sum_{m=0}^{\infty}\sum_{n=0}^{\infty}  P_{\rm m}(m,n)\,z^m\, 
\lambda^n=G(1,z,\lambda)=\frac{1}{1-\lambda}\, \frac{z_0-1}{z_0-z}.
\label{gen-max}
\end{equation}
The normalization condition $G(1,1,\lambda)=(1-\lambda)^{-1}$ is
immediately checked. It is easy to invert $G(1,z,\lambda)$ with
respect to $z$ exactly, which gives
\begin{equation}
\sum_{n=0}^\infty P_\mathrm{m}(m,n) \lambda^n =\frac{z_0-1}{1-\lambda} \,
\frac{1}{z_0^{m+1}} \;,
\label{P_m} 
\end{equation}
where $P_\mathrm{m}(m,n)$ is the probability of having the maximum at
$m$ in $n$ steps.

On the other hand, substituting $F(z,\lambda)$ in \eref{Gsz-2} yields
the full generating function as
\begin{equation}
G(s,z,\lambda)=\frac{2s_+}{a} \,
\frac{1 - r (1-s s_{-}) }{(s_+-s) (z_0-z)}, 
\end{equation}
where we have used $s_+s_-=1$ and 
\begin{equation*}
z_0-z_1(s)= (s-s_-) s^{-1} s_+
\bigl[ 1- r(1-s s_-)\bigr] (1-\lambda) (1-a)^{-1}.
\end{equation*}
Using $z_0=z_1(s_-)$, it can be shown that
\begin{equation}
(z_0-1) (1-s_-)=2a^{-1}(1-\lambda) [1 -r (1-s_{-})].
\label{useful-relation}
\end{equation}
Therefore, the normalization condition
$G(1,1,\lambda)=(1-\lambda)^{-1}$ is checked from the above
expression.  Finally, it is useful to rewrite the above expression as
\begin{equation}
G(s,z,\lambda)=\frac{2}{a} \,
\left[\frac{s_{+}}{s_+-s} -r \right]\,\frac{1}{z_0-z} , 
\label{mainresult}
\end{equation}
where we recall that Eqs.~\eqref{a}, \eqref{s_pm} and
\begin{equation}
\label{z0}
z_0=
\frac{(1-\lambda)}{(1-a)}\left[-rs_{-}+\frac{(1-r)}{s_{-}} + 2r+
\frac{r\lambda}{(1-\lambda)}\right],
\end{equation}
using \eref{z1} in $z_0=z_1(s_{-})$.  The explicit expression of the
generating function $G(s,z,\lambda)$ in Eq. (\ref{mainresult}) is the
central result of this paper and is valid for arbitrary $0\le r\le 1$.

Let us first check a few immediate consequences of this result in
Eq. (\ref{mainresult}). One can easily invert Eq. (\ref{mainresult})
with respect to $s$ and $z$ to give
\begin{equation}
\sum_{n=0}^\infty Q(y,m,n) \lambda^n = \frac{2}{a}\, \left[s_+^{-y}
  -r\delta_{y,0}\right] \, z_0^{-(m+1)}.
\label{Q(y,m,n)}
\end{equation}
Now, using $P(x,m,n)=Q(m-x,m,n)$ we get
\begin{equation}
\sum_{n=0}^\infty P(x,m,n) \lambda^n = \frac{2}{a z_0}\,\left[  s_+^{x} \,
(s_+ z_0)^{-m} -r\delta_{x,m} z_0^{-m}\right].
\label{Pxm}
\end{equation}

From Eq. (\ref{Pxm}), one can derive the marginal distribution of the 
position
by summing over $m$. Note that for positive $x$, the sum over $m$ goes
from $x$ to $\infty$. In contrast, for negative $x$, the sum goes from zero
to $\infty$.  This yields
\begin{align}
\label{P_x}
\sum_{n=0}^\infty  P_\mathrm{x}(x,n) &\lambda^n = \frac{2}{a (s_+z_0 -1)}\notag\\
&\times
\begin{cases}
\bigl[(1-r) s_{+} z_0 +r\bigr] z_0^{-(x+1)} & \text{for $x \ge 0$},\\[2mm]
s_+^{(x+1)}  & \text{for $x < 0$}.
\end{cases} 
\end{align}

For completeness, we also compute the generating function with respect
to both $x$ and $n$, which gives
\begin{equation}
\sum_{x=-\infty}^\infty \sum_{n=0}^\infty P_\mathrm{x}(x,n)\, s^x\, \lambda^n =
\frac{2}{a}\, \frac{s-r(s-s_-)}{ (z_0-s) (s-s_-)}.
\label{Pxn-GF}
\end{equation}
Setting $s=1$ above and using \eref{useful-relation}, it is easy to
verify the normalization condition
\begin{equation}
\sum_{x=-\infty}^\infty \sum_{n=0}^\infty P_\mathrm{x}(x,n) \lambda^n
= (1-\lambda)^{-1}.
\end{equation}

As mentioned above, the expression \eref{mainresult} for the full
generating function $G(s,z,\lambda)$, valid for arbitrary 
reset probability $0\le r\le 1$,
is the main central 
result
of our paper. Several asymptotic results for the statistics of the
two random variables $m$ and $x$ can then be derived by analyzing
this expression in \eref{mainresult} in different limits, which we
present in the next few sections. 

\section{The $r=0$ case}
\label{r=0 case}

Let us first check the case without resetting, i.e., we set $r=0$.
This is the standard one dimensional random walk and the results
for the maximum and the position of the walker are well known.
However, we reproduce it here as a check as well as for the sake
of completeness. 

For $r=0$, $a=\lambda$, and 
\begin{equation}
z_0=s_+=\frac{1+\sqrt{1-\lambda^2}}{\lambda}.
\end{equation}
Putting $\lambda=e^{-p}$, and taking $p\to 0$
limit, we have $z_0 \sim1+\sqrt{2p}$, $(z_0-1)/(1-\lambda) \sim
\sqrt{(2/p)}$ and $z_0^m \to \exp(m\sqrt{2p})$
as $m\to\infty$ keeping $m\sqrt{p}$ fixed.  Therefore,
in this limit, from~\eref{P_m}
\begin{equation}
\sum_n P_\mathrm{m}(m,n) e^{-p n}\, 
\sim \frac{\sqrt{2}}{\sqrt{p}}\,
\exp\Bigl(-m\sqrt{2p}\Bigr). 
\end{equation}
Inverting the Laplace transform gives, 
\begin{equation}
P_\mathrm{m}(m,n) \sim \sqrt{\frac{2}{\pi n}} \exp\left( -\frac{m^2}{2
  n}\right), ~~~\mathrm{where}~m \ge 0. 
\label{pmmr0}
\end{equation}
From this distribution, it is easy to compute all the moments for large 
$n$. For instance, the mean and the variance of the maximum
grow asymptotically as
\begin{equation}
\langle m(n)\rangle \simeq  \sqrt{\frac{2\,n}{\pi}} \quad\text{and}\quad
\sigma_m^2 \simeq  \left(1-\frac{2}{\pi}\right)\, n ~, 
\end{equation}
as stated respectively in Eqs. (\ref{r0meanm}) and (\ref{r0mdiff}). 

Similarly, from \eref{P_x}, we get,
\begin{equation}
\sum_n P_\mathrm{x}(x,n) e^{-p n}\, \sim \frac{1}{\sqrt{2\,p}}\,
\exp\Bigl(-|x|\sqrt{2\,p}\Bigr), 
\end{equation}
which, after Laplace inversion, gives the expected Gaussian distribution (since
a random walk, for large $n$, converges to a Brownian motion)
\begin{equation}
P_\mathrm{x}(x,n) \sim \frac{1}{\sqrt{2\pi n}} \exp\left( -\frac{x^2}{2
  n}\right).
\end{equation}
Thus the mean and the variance of the position behave as
\begin{equation}
\langle x(n)\rangle = 0 \quad\text{and}\quad
\sigma_x^2 \simeq  n ~,
\end{equation}
as stated respectively in Eqs. (\ref{r0meanx}) and (\ref{r0diffcx}).

For the joint distribution of the maximum and the position, we get from 
\eref{Pxm} 
\begin{equation}
 \sum_{n=0}^\infty P(x,m,n) e^{-p n} \sim 2 
\exp\Bigl[-(2m-x)\sqrt{2\,p}\Bigr],
\end{equation}
which, after Laplace inversion, gives
\begin{equation}
P(x,m,n) \sim \frac{(2 m-x)}{n^{3/2}}\sqrt{\frac{2}{\pi }}
\exp\left[{-\frac{(2 m-x)^2}{2 n}}\right],
\label{P(x,m,n)}
\end{equation}
where $x\le m$ and $m\ge 0$.

Let us also verify \eref{Pxn-GF} using the exact result of the random
walk 
\begin{equation}
P_\mathrm{x}(x,n)={n\choose\frac{n+x}{2}} 2^{-n}, ~~\mathrm{when}~ n+x
~\mbox{is even},
\end{equation}
and zero otherwise. Therefore,
\begin{equation}
\sum_{x=-n}^{n} P_\mathrm{x}(x,n) s^x= \sum_{m=0}^n P(2m-n, n) s^{2m-n}
=\left[\frac{1}{2} \left(s+\frac{1}{s}\right)\right]^n.\notag
\end{equation}
Consequently, 
\begin{align}
\sum_{n=0}^\infty \lambda^n \sum_{x=-n}^{n} P_\mathrm{x}(x,n) s^x &=
\left[1-\frac{\lambda}{2} \left(s+\frac{1}{s}\right)
  \right]^{-1}\notag \\
&=\frac{2s}{\lambda} \, \frac{1}{(s_+-s) (s-s_-)}\, ,
\end{align}
which is same as \eref{Pxn-GF} for $r=0$.

Thus, for the case $r=0$, when there is no resetting to the maximum,
we have recovered the results for the usual random walk.

\section{The $r\to 1$ limit}
\label{r->1 limit}

In the other extreme limit of $r\to 1$, we 
have
\begin{math}
z_0 =
(2-\lambda^2)/\lambda.
\end{math}
Therefore, from \eref{P_m}
\begin{equation}
\sum_{n=0}^\infty P_\mathrm{m}(m,n) \lambda^n =\frac{2+\lambda}{2-\lambda^2} \,
\left[\frac{\lambda}{2-\lambda^2}\right]^{m}.
\label{pmr1}
\end{equation}
From the series expansion of the above expression with respect to
$\lambda$, it is easily verified that
$P_\mathrm{m}(m,0)=\delta_{m,0}$. Moreover,
\begin{equation}
P_\mathrm{m}(0,n)= 2^{-\frac{(n+1)}{2}}
\left[\frac{1-(-1)^n}{2}+\sqrt{2}\,\frac{1+(-1)^n}{2} \right], 
\end{equation}
is the probability that the walker has not crossed the origin up to
the step $n$. This is the case where the walker goes to the site
$x=-1$ (with probability $1/2$) and comes back to the origin (with
probability one) at alternate time steps.  In general (for $r\to 1$), 
\begin{equation}
 P_\mathrm{m}(m,n)  =\frac{1}{2\pi i}\oint_0 \frac{d\lambda}{\lambda^{n+1}}\, 
\frac{2+\lambda}{2-\lambda^2} \,
\left[\frac{\lambda}{2-\lambda^2}\right]^{m}.
\end{equation}
For large $m$ and $n$, with $m=w n$, the integral can be evaluated
using saddle-point approximation, which to the leading order gives
\begin{equation}
P_\mathrm{m} (m=wn, n) \sim \exp[-n S(w)], 
\end{equation}
where $S(w)\equiv  S(w,\lambda^*)$ with
\begin{equation}
S(w,\lambda) =\ln\lambda
-w \ln\left[\frac{\lambda}{2-\lambda^2}\right] .
\end{equation}
The saddle point $\lambda^*$ is obtained by solving the equation
$\partial_\lambda S(w,\lambda)|_{\lambda^*}=0$, as
\begin{math}
[\lambda^*(w)]^2= 2(1-w)/(1+w).
\end{math}
Substituting this in the above equation we obtain the large deviation
function as
\begin{equation}
S(w)\equiv S(w,\lambda^*) =   \frac{(1-w)}{2} \ln \frac{2(1-w)}{(1+w)} 
+  w \ln \frac{4w}{(1+w)}.
\label{S-LDFr1}
\end{equation}
This large deviation function has a maximum at $w=1/3$, and near this,
one gets $S(w)=(27/16) (w-1/3)^2$, which implies the
Gaussian form
\begin{equation}
P_\mathrm{m}(m,n) \sim \exp\left(-\frac{(m-n/3)^2}{2 D n}\right),
~~\mathrm{with}~~D=\frac{8}{27}. 
\end{equation}
In fact, \eref{pmr1} can be inverted exactly, which gives 
\begin{equation}
P_\mathrm{m}(m,n)=
\begin{cases}\displaystyle
2^{-\frac{(n+m)}{2}} 
{\frac{n+m}{2}\choose m}
&\text{if $(n-m)$ is even},\\[5mm]
\displaystyle
2^{-\frac{(n+m+1)}{2}} 
{\frac{n+m-1}{2}\choose m}
&\text{if $(n-m)$ is odd}.
\end{cases}
\label{p(m,n) for r->1}
\end{equation}
For large $n$ and $m$, using the Stirling's approximation in
\eref{p(m,n) for r->1}, one can recover the large deviation function
given by \eref{S-LDFr1}.

Let us now look at the probability distribution of the position using
\eref{P_x}. For $r\to 1$ we have $a\to 0$ and $s_{+}\to
\infty$ with $as_{+}\to 2$. Therefore, from \eref{P_x}, it is
clear that for negative $x$, we get nonzero probability only for
$x=-1$, which reads
\begin{equation}
\sum_{n=0}^\infty P_\mathrm{x}(-1,n)\lambda^n= \frac{1}{z_0}
=\frac{\lambda}{2-\lambda^2}.
\end{equation}
Inverting this with respect to $\lambda$ gives
\begin{equation}
P_\mathrm{x}(-1,n)=\frac{1-(-1)^n}{2}
~2^{-\frac{n+1}{2}}.
\end{equation}
This result can be understood, as this is the case where the walker
goes to the site $x=-1$ (with probability $1/2$) at odd time steps and
comes back to the origin (with probability one) at even time
steps. The also implies that $P(-1,n)$ is nonzero only when the
maximum remains zero. Indeed from \eref{Pxm}, for $r\to 1$ we
get $\sum_n P(-1,0,n)\lambda^n =z_0^{-1}=\sum_n P_\mathrm{x}(-1,n)\lambda^n$.
For $x\ge 0$, \eref{P_x} we get
\begin{equation}
\sum_{n=0}^\infty P_\mathrm{x}(x,n) \lambda^n =\frac{4-\lambda^2}{(2-\lambda^2)^2} \,
\left[\frac{\lambda}{2-\lambda^2}\right]^{x}.
\end{equation}
Although the exact form differs from \eref{pmr1}, it is evident that
$P_\mathrm{x}(x,n)$ and $P_\mathrm{m}(m,n)$ have the same large deviation function,
i.e., 
\begin{equation}
P_\mathrm{x} (x=wn, n) \sim \exp[-n S(w)],
\end{equation}
with $S(w)$  given by \eref{S-LDFr1}.

\section{The case $0 < r <1$}
\label{general resetting}

For general resetting probability $0<r<1$, it is a bit cumbersome to
find the exact large deviation functions associated with the
probabilities $P_\mathrm{m}(m,n)$ and $P_\mathrm{x}(x,n)$. However, one
expects (as in the cases of $r=0, 1$) the typical fluctuations near
the mean to be governed by Gaussian distributions. The generating
functions for the mean and the variance for the maximum and the
current position are obtained from the generating functions of their
distributions,  given
by \eref{gen-max} and \eref{Pxn-GF} respectively, simply by taking
derivatives. From their respective generating functions, we then derive,
for arbitrary but fixed $0<r<1$,
the asymptotic behavior of the mean and variance for large $n$ for both
the maximum and the position. Finally, the asymptotic behavior of the
the full probability distribution of the difference variable $y(n)=m(n)-x(n)$
is derived. These derivations are outlined in the next three subsections.

\subsection{Statistics of the maximum $m(n)$}

From \eref{gen-max}, we get the generating functions of the first two
moments of the maximum as
\begin{equation}
\sum_{n=0}^\infty \langle m(n) \rangle \lambda^n
=\frac{\partial}{\partial z} G(1,z,\lambda) \Big|_{z=1}
= \frac{1}{1-\lambda}\, \frac{1}{z_0 -1}, 
\label{moment-1}
\end{equation}
and
\begin{align}
\sum_{n=0}^\infty \langle m^2(n) \rangle \lambda^n
&=\frac{\partial}{\partial z} z \frac{\partial}{\partial z}
G(1,z,\lambda) \Big|_{z=1} \notag\\
&= \frac{1}{1-\lambda}\,\left[\frac{2}{(z_0-1)^2} +  \frac{1}{z_0
    -1}\right], 
\label{moment-2}
\end{align}
respectively. Using the expression of $z_0=z_1(s_{-})$ from Eq.~(\ref{z1}),
it is easy to check that  
\begin{equation}
\label{1/(z0-1)}
\frac{1}{z_0 -1} = \frac{C(r,\lambda)}{1-\lambda},
\end{equation}
where
\begin{equation}
C(r,\lambda)=\frac{a(1-a)}{(1-a) (1-2 r) +\sqrt{1-a^2}},
\label{C(r,lambda)}
\end{equation}
with $a=(1-r) \lambda$.  Inverting Eqs.~(\ref{moment-1}) and
(\ref{moment-2}) with respect to $\lambda$ using Cauchy's formula, we
get
\begin{equation}
\langle m(n) \rangle = \frac{1}{2\pi i}\oint_0
\frac{d\lambda}{\lambda^{n+1}} \frac{C(r,\lambda)}{(1-\lambda)^2}
\label{m1_cauchy}
\end{equation}
and
\begin{equation}
\langle m^2(n) \rangle = \frac{1}{2\pi i}\oint_0
\frac{d\lambda}{\lambda^{n+1}} \left[\frac{2
    C^2(r,\lambda)}{(1-\lambda)^3} + \frac{C(r,\lambda)}{(1-\lambda)^2}
  \right]
\label{m2_cauchy}
\end{equation}
respectively, where the integral is along a counterclockwise closed
contour around the origin in the complex $\lambda$ plane. 

A hint that $r=0$ is a special case and is different from the $r>0$ case
can be already seen at this level. For $r=0$, $a=\lambda$ and 
from Eq. (\ref{C(r,lambda)}) one gets 
$C(0,\lambda)=(1-\lambda) \lambda \bigl[(1-\lambda)
  +\sqrt{1-\lambda^2}\bigr]^{-1}$. Therefore, for the $r=0$ case, the
powers of $(1-\lambda)$ in the denominators of the above expressions,
reduce by one, and hence this case must be treated separately from the
$r>0$ case. In the
following we will only consider the case $r>0$.

Evaluating the contour integrals in Eqs. (\ref{m1_cauchy}) and (\ref{m2_cauchy})
explicitly for all $n$ looks cumbersome. However, the asymptotic behavior for
large $n$ can be derived by separating out the contributions
to the contour integrals arising from the pole at $\lambda=1$.
Note that $C(r,\lambda)$ also has two branch points at $\lambda=\pm
(1-r)^{-1}$. Therefore, $\oint_0 =-\oint_1 + $ [contributions from the
  integrals around the branch cuts from $-\infty$ to $-(1-r)^{-1}$ and
  from $(1-r)^{-1}$ to $\infty$].  Using the residue theorem and
computing the residue at $\lambda=1$ we get
\begin{align}
\label{<m>}
&\frac{1}{2\pi i}\oint_0
\frac{d\lambda}{\lambda^{n+1}} \frac{C(r,\lambda)}{(1-\lambda)^2} =
C_0(r) (n+1) + C_1(r) \notag\\
&\qquad\qquad\qquad\qquad
+ [\text{branch cuts contributions}],
\intertext{and}
\label{<m^2>}
&\frac{1}{2\pi i}\oint_0
\frac{d\lambda}{\lambda^{n+1}} \frac{2
    C^2(r,\lambda)}{(1-\lambda)^3} =
 (n+1) (n+2) C_0^2(r) \notag \\
&\qquad\qquad
+ 4 (n+1) C_0(r) C_1(r) 
+2 C_1^2(r) + 2 C_0(r) C_2(r) \notag\\
&\qquad\qquad\qquad\qquad
+ [\text{branch cuts contributions}],
\end{align}
where 
\begin{equation}
C_n(r)=(-1)^n \frac{\partial^n}{\partial \lambda^n}
C(r,\lambda)\Big|_{\lambda=1} = \frac{\partial^n}{\partial \lambda^n}
C(r,1-\lambda)\Big|_{\lambda=0}. 
\label{Cn}
\end{equation}
These coefficients can be calculated explicitly using Eq. (\ref{C(r,lambda)}).
For example, the first two coefficients are given explicitly as
\begin{align}
C_0(r)& = \frac{r(1-r)}{r(1-2r)+\sqrt{2r-r^2}} \label{C0r} \\
\intertext{and}
C_1(r) & =  
\frac{r(1-r)\left[1-3r + r^2+r(2r-1)\sqrt{2r-r^2}\right]}{\sqrt{2r-r^2}
\left(r-2r^2+\sqrt{2r-r^2}\right)^2}. \label{C1r}
\end{align}
One can also show that the branch cuts contributions in Eqs. (\ref{<m>})
and (\ref{<m^2>}) go to 
zero exponentially fast as 
$n\rightarrow \infty$.

Using the results from Eqs. (\ref{<m>}) and (\ref{<m^2>}) in
Eqs. (\ref{m1_cauchy}) and (\ref{m2_cauchy}), we then obtain,
for large $n$, the mean
\begin{equation}
\langle m(n) \rangle  \simeq v(r)\, n,~\mbox{with}~ 
v(r)=C_0(r)=\frac{r(1-r)}{r(1-2r)+\sqrt{2r-r^2}}.
\label{meanm1}
\end{equation}
The speed $v(r)$, as a function of $r$, is plotted in \fref{v&D}.
Similarly, the variance grows linearly for large $n$ 
\begin{align}
\sigma_m^2=\langle m^2 \rangle- \langle m \rangle^2 
\simeq  D_m(r)\, n,\\
 \intertext{with}
D_m(r)=D(r)\equiv C_0^2(r) + 2 C_0(r)\, 
C_1(r) 
+C_0(r).
\end{align}
Using $C_0(r)$ and $C_1(r)$ from Eqs. (\ref{C0r}) and (\ref{C1r})
gives the explicit expression for the diffusion coefficient $D(r)$,
for $r>0$, as given by \eref{D(r)}. A plot of $D(r)$ vs. $r$ is
provided in \fref{v&D}.


\subsection{Statistics of the position $x(n)$}

Similarly, to compute the mean and variance of the position $x(n)$, we take
derivatives with respect to $s$ in \eref{Pxn-GF} at $s=1$ to
get
\begin{equation}
\sum_{n=0}^\infty \langle x(n) \rangle \lambda^n  = \frac{2}{a} \biggl[ 
\frac{1-r(1-s_-)}{\left(1-s_-\right) \left(z_0-1\right)^2} 
-\frac{s_-}{\left(1-s_-\right)^2 \left(z_0-1\right)}
\biggr] 
\label{m1x}
\end{equation}
and
\begin{align}
&\sum_{n=0}^\infty \langle x^2(n) \rangle \lambda^n = \frac{2}{a} \biggl[ 
\frac{2 \left[1-r \left(1-s_-\right)\right]}{\left(1-s_-\right)
  \left(z_0-1\right)^3} \notag\\
&\qquad
 + \frac{1-r \left(1-s_-\right)^2-3 s_-}{\left(1-s_-\right)^2
  \left(z_0-1\right)^2}
+\frac{s_- \left(1+s_-\right)}{\left(1-s_-\right)^3 \left(z_0-1\right)}
\biggr],  
\label{m2x}
\end{align}
where we recall Eqs.~\eqref{a}, \eqref{s_pm} and \eqref{z0}.

It is useful to define the following quantities
\begin{align}
A_1(r,\lambda)&= \frac{2[1-r(1-s_-)]}{a(1-s_-)},\\
A_2(r,\lambda)&= \frac{2s_-}{a(1-s_-)^2},\\
A_3(r,\lambda)&=\frac{2[1-r \left(1-s_-\right)^2-3
    s_-]}{a\left(1-s_-\right)^2},\\
A_4(r,\lambda)&=\frac{2 s_- \left(1+s_-\right)}{a \left(1-s_-\right)^3}.
\end{align}
We note that $A_1(r, \lambda) C(r,\lambda)=1$ and $[2A_2(r,\lambda) +
  A_3(r,\lambda) ] C(r,\lambda)=1$. Using these definitions and
\eref{1/(z0-1)}, we invert the above generating function with respect
to $\lambda$ and get
\begin{align}
\langle x(n) \rangle &= \frac{1}{2\pi i}\oint_0
\frac{d\lambda}{\lambda^{n+1}} \left[
  \frac{C(r,\lambda)}{(1-\lambda)^2} -\frac{A_2(r,\lambda)
    C(r,\lambda)}{(1-\lambda)} \right], \\  
\langle x^2(n)
\rangle &= \frac{1}{2\pi i}\oint_0 \frac{d\lambda}{\lambda^{n+1}}
\biggl[\frac{2 C^2(r,\lambda)}{(1-\lambda)^3} + \frac{A_3(r,\lambda)
    C^2(r,\lambda)}{(1-\lambda)^2} \notag\\
&\qquad\qquad\qquad\qquad\qquad
+ \frac{A_4(r,\lambda)
    C(r,\lambda)}{(1-\lambda)} \biggr].
\end{align}
Finally, evaluating these integrals using residue theorem, we get
\begin{align}
\langle x(n) \rangle &= (n+1) C_0(r) + C_1(r)
-A_{2,0}(r) C_0(r) +\dots, \\
\intertext{and}
\langle x^2(n) \rangle &= (n+1)(n+2) C_{0}^2\notag\\
&+(n+1) \left[ 4 C_0 (r) C_1(r) 
+ A_{3,0}(r) C_0^2(r)
  \right]\nonumber \\
&+2 C_1^2(r) + 2 C_0(r) C_2(r)
 + A_{3,1}(r) C_0^2(r) \notag\\
&+ 2 A_{3,0} (r)
C_0(r) C_1(r) 
+A_{4,0}(r) C_0(r) +\dots ,
\end{align}
where $C_n(r)$ is defined by \eref{Cn} and  
\begin{equation}
A_{l,n}(r)=(-1)^n \frac{\partial^n}{\partial \lambda^n}
A_l(r,\lambda)\Big|_{\lambda=1} = \frac{\partial^n}{\partial \lambda^n}
A_l(r,1-\lambda)\Big|_{\lambda=0}. 
\end{equation}
 
Finally, collecting the leading term for large $n$, we obtain the
following asymptotic results for the mean and the variance.
For large $n$, and for $r>0$, the mean
\begin{equation}
\langle x(n) \rangle \sim v(r)\, n, 
\label{meanx.1}
\end{equation}
where the speed $v(r)=C_0(r)$ turns out to be exactly the same as in
case of maximum, with explicit expression given in Eq. (\ref{meanm1}). 
Similarly, the variance behaves, asymptotically for large $n$ and
for all $r>0$, as 
\begin{align}
&\sigma_x^2=\langle x^2(n) \rangle- \langle x(n) \rangle^2
\simeq  D_x(r)\, n, \nonumber \\
&\text{with}~~ D_x(r)=D(r)\equiv C_0^2(r) + 2 C_0(r)\,
C_1(r)
+C_0(r).
\label{varx.1}
\end{align}
Hence, the first and the second moments of the two variables
$m(n)$ and $x(n)$ grow identically for large $n$.
It is then natural to ask how the difference variable
$y(n)=m(n)-x(n)$ is distributed and we address this in the
next subsection


\subsection{Asymptotic distribution of the difference $y(n)=m(n)-x(n)$}

Let $Q_\mathrm{y}(y,n)$ be the probability that the location
at the $n$-th step is at a distance $y$ away from the global
maximum. Clearly, $Q_\mathrm{y}(y,n)=\sum_{m=0}^\infty
Q(y,m,n)$. Therefore, from \eref{Q(y,m,n)} we get
\begin{equation}
\sum_{n=0}^\infty Q_\mathrm{y}(y,n)\lambda^n=
\frac{2}{a}\, \left[s_+^{-y}
  -r\delta_{y,0}\right] \, \frac{1}{z_0-1}.
\label{Q_y-gen}
\end{equation}

Let us first look at the value $y=0$, for which Eq. (\ref{Q_y-gen})
reads
\begin{equation}
\sum_{n=0}^\infty Q_\mathrm{y}(0,n)\lambda^n=
\frac{2}{\lambda}\,\frac{1}{z_0-1}=\frac{2 C(r,\lambda)}{\lambda(1-\lambda)}
\label{Q_0-gen}
\end{equation}
where we have used Eq. (\ref{1/(z0-1)}). To extract the large $n$ behavior
of $Q_\mathrm{y}(0,n)$, we need to investigate the right hand side of 
\eref{Q_0-gen}
in the limit $\lambda\to 1$. As $\lambda\to 1$, $C(r,\lambda)\to C(r,1)=C_0(r)$
given in Eq. (\ref{C0r}). Hence, the right hand side of Eq.~(\ref{Q_0-gen})
behaves as $2C_0(r)/(1-\lambda)$ as $\lambda\to 1$. This clearly
indicates that
$Q_\mathrm{y}(0,n)$ becomes independent of
$n$ in the limit $n\to\infty$, given by
\begin{equation}
Q_\mathrm{y}(0,n\to\infty)\to 2 C_0(r). 
\label{Q_y(0)}
\end{equation}

For $y>0$, inverting the generating function \eref{Q_y-gen} we get
\begin{equation}
Q_\mathrm{y}(y,n)=\frac{1}{2\pi i}\oint_0
\frac{d\lambda}{\lambda^{n+1}} \frac{2}{a(z_0-1)} s_+^{-y}.
\end{equation}
Now, using explicit expressions, and changing the integral over
$\lambda$ to that over $a=(1-r)\lambda$, we get
\begin{align}
Q_\mathrm{y} (y,n)=\frac{1}{2\pi i} \oint_0
&\frac{da}{(a_0-a)}
\left[\frac{2(1-a)}{(1-a) (2a_0-1) +\sqrt{1-a^2}}\right]\notag\\
&\times\left(\frac{a_0}{a}\right)^{n+1}
\left[\frac{a}{1+\sqrt{1-a^2}}\right]^y,
\end{align}
where $a_0=1-r$. The integrand has a simple pole at $a=a_0$ and branch
points at $a=\pm 1$. The contour of integration around zero can be
split into two vertical contours: one that goes from $+i\infty$ to
$-i\infty$ through the left of the origin and another that goes from
$-i\infty$ to $+i\infty$ through the right of the origin. The left
contour subsequently can be wrapped around the branch cut from $a=-1$
to $-\infty$. The contribution from this contour is subdominant and
the main contribution comes from the contour on the right for large
$n$. It is useful to express the integral as
\begin{equation}
Q_\mathrm{y}(y=wn,n)  \approx \frac{1}{2\pi i}
\int_{(0_{+})-i\infty}^{(0_{+})+i\infty} da\frac{g(a)}{(a_0-a)}
\exp\bigl[-n H(w,a)\bigr],
\label{Q_y large.1}
\end{equation}
so that for large $n$ we can use the saddle point approximation method.
Here
\begin{equation}
H(w,a)=\ln(a/a_0)
-w\ln \left[\frac{a}{1+\sqrt{1-a^2}}\right],
\label{H(w,a)}
\end{equation}
and
\begin{equation}
g(a)=
\frac{2(1-a)}{(1-a) (2a_0-1) +\sqrt{1-a^2}}.
\end{equation}
The saddle point $a^*$ is obtained by solving the condition $\partial_a
H(w,a)|_{a^*}=0$, which gives $a^*=\sqrt{1-w^2}$. Note that the integrand of
\eref{Q_y large.1} has a simple pole at $a=a_0=1-r$. 
For $w< w^*=\sqrt{r(2-r)}$, we have $a^* > a_0$, and therefore, the
contribution to the above integral comes from both the pole and the
saddle point. However, the contribution from the pole is larger than
that from the saddle point. Therefore, for large $n$,
\begin{math}
Q_\mathrm{y}(y=wn,n)\sim\exp[-n H_1(w)],
\end{math}
where the large deviation function is given by
\begin{equation}
H_1(w)\equiv H(w,a_0)=w\ln\left[\frac{1+\sqrt{r(2-r)}}{1-r}\right].
\label{pole form}
\end{equation}
On the other hand, for $w > w^*$, we have $a^* < a_0$.
Therefore, evaluating the above integral using saddle-point
approximation gives
\begin{math}
Q_\mathrm{y}(y=wn,n)\sim\exp[-n H_2(w)]
\end{math}
where the large deviation function is given by
\begin{equation}
H_2(w)\equiv H(w,a^*)=\frac{w}{2}\ln\frac{1+w}{1-w} + \ln
  \frac{\sqrt{1-w^2}}{1-r} . 
\label{saddle form}
\end{equation}
Finally, combining these two regimes we obtain the large deviation
behavior
\begin{equation}
Q_\mathrm{y}(y=wn,n)  \sim
\exp\bigl[-n H(w)\bigr],
\label{Q_y large.2}
\end{equation}
with the rate function given by
\begin{equation}
H(w)=
\begin{cases}
\displaystyle
H_1(w)=
w\ln\left[\frac{1+\sqrt{r(2-r)}}{1-r}\right] &\text{for $w<w^*$},\\[5mm]
\displaystyle
H_2(w)=\frac{w}{2}\ln\frac{1+w}{1-w} + \ln
  \frac{\sqrt{1-w^2}}{1-r}   &\text{for $w> w^*$},
\end{cases}
\label{H(w).1}
\end{equation}   
with $w^*=\sqrt{r(2-r)}$.  

\begin{figure}
\centerline{\includegraphics[width=0.95\hsize]{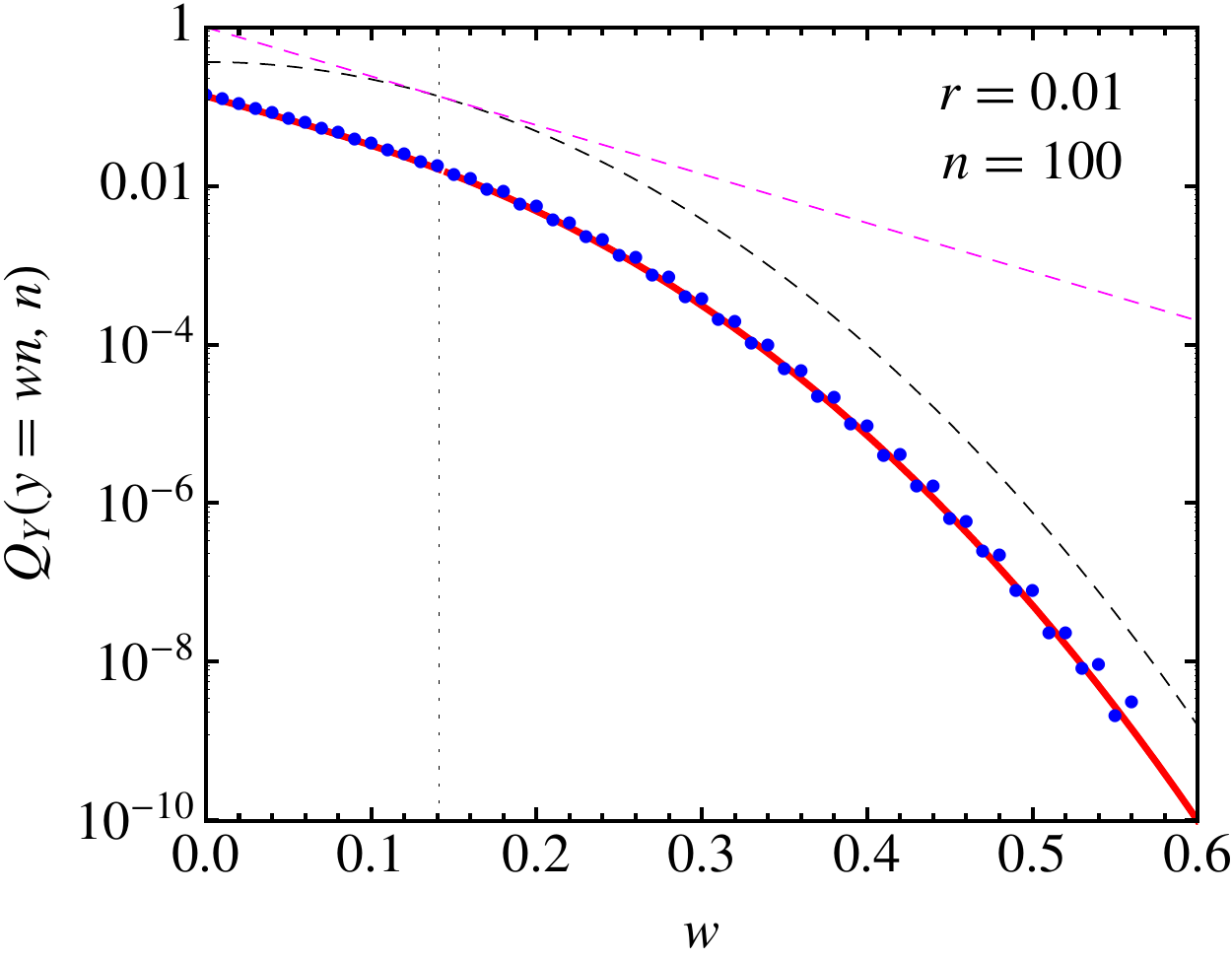}}
\caption{\label{PDF}(Color online) $Q(y,n)$ against the scaled
  variable $w=y/n$ for $n=100$ and $r=0.01$. The (blue) points are
  obtained from numerical simulation. The solid (red) line represents
  the asymptotic form given by \eref{PDF-asymptotic}. The dashed lines
  (magenta and black) plot the large deviation forms
  $Q_\mathrm{y}(y=wn,n)\sim\exp[-nH_{1,2}(w)]$ with the large
  deviation functions $H_1(w)$ and $H_2(w)$ given by \eref{pole form}
  and \eref{saddle form} respectively. The vertical dotted line marks
  the position of $w^*\approx 0.141$.}
\end{figure}

As discussed in Section 2, this result
indicates that for a given large $n$,
$Q_{\rm y}(y,n)$ becomes independent of $n$ for
$y< w^* n$ and is still $n$-dependent for $y>w^*\, n$, signalling
a {\it dynamical phase transition}. The rate function $H(w)$, plotted
in \fref{PDF}, is weakly singular at the critical point $w=w^*$ where
both $H(w)$ and $H'(w)$ are continuous, but
the second derivative $H''(w)$ is discontinuous: $H''(w\to {w^*}^{-})=0$,
while $H''(w\to {w^*}^{+})= 1/(1-r)^2$. Such a second order dynamical
phase transition was also observed recently in the
time evolution of the 
distribution of 
position of a Brownian motion in one dimension
with resetting to its initial position~\cite{Majumdar:2015}.


We finish this subsection by making a couple of interesting observations.
In the limit $r\to 0$ we have $w^*=0$. Therefore, the large deviation
form is given by $H_2(w)$  with $r=0$. Expanding in Taylor
series, for small $w$ we get $H_2(w) \simeq w^2/2$, which gives the
Gaussian form $Q_\mathrm{y}(y,n) \sim \exp [-y^2/(2n)]$. This is
consistent with the result obtained by substituting $x=m-y$ in
\eref{P(x,m,n)} and integrating over $m$ from $0$ to $\infty$.
On the other hand, in the limit $r\to 1$, we have $w^*=1$. Therefore,
the large deviation form is given by $H_1(w)$ with $r=1$. We
note that for $r=1$, the large deviation function is $-\infty$, except
for the case $w=0$. This is because, $y$ can take only two values,
namely, $0$ and $1$, for the case $r=1$. Using the result of
\eref{Q_y(0)}, we get $Q_\mathrm{y}(0,n\to\infty)\to 2C_0(1)=2/3$. For
$y=1$, \eref{Q_y-gen} gives 
\begin{equation}
\sum_{n=0}^\infty Q_\mathrm{y}(1,n)\lambda^n=
\frac{2}{(a s_+) (z_0-1)}.
\end{equation}
Now, in the limit $r\to 1$, we get $as_+\to 2$ and $(z_0-1)\to
\lambda^{-1}(1-\lambda) (2+\lambda)$. Using these, and inverting the
above equation we get
\begin{equation}
Q_\mathrm{y}(1,n) =\frac{1}{3}\left[1-\left(-\frac{1}{2}\right)^n\right]
\to \frac{1}{3}\quad\mbox{as}~ n\to\infty.
\end{equation}

Finally, following the method used in Ref.~\cite{Sabhapandit:2012fm}
in a different context, we 
can also write down a more
complete asymptotic form of $Q(y,n)$  for large $n$ as
\begin{widetext}
\begin{align}
Q_\mathrm{y}(y=wn,n) \approx  \frac{ e^{-n
H_2(w)}}{2\sqrt{\pi n}}\left[K(w)
-\frac{\sgn{w^*-w}\,g_{-1}}{\sqrt{H_2(w)-H_1(w)}}\right]
+  e^{-n H_1(w)}g_{-1} \left[
 \frac{\sgn{w^*-w}}{2}\, \erfc \bigl(\sqrt{n[H_2(w)-H_1(w)]}\,\bigr)
 -\theta(w^*-w)
\right],
\label{PDF-asymptotic}
\end{align}
\end{widetext}
where $g_{-1}=-g(a_0)$ and
\begin{math}
K(w)=\sqrt{2}\, w\, g(a^*)/(a_0-a^*).
\end{math}
This pre-asymptotic form is particularly useful to compare to the 
results of simulation. Indeed,
\Fref{PDF} compares this form with the
numerical simulation results. The agreement is excellent. 

\section{The limit $r\to 0$ and the associated scaling functions}
\label{scaling-limit}

From the results presented in \Sref{r=0 case} for $r=0$ and 
in \Sref{general resetting} for $r>0$, we see that the limit $r\to 0$ (after
taking the large $n$ limit) is not the same as $r=0$, for the statistics
of both the maximum and the position. In other words, the two limits
$\lim_{r\to 0}$ and $\lim_{n\to \infty}$ do not commute. This indicates that
$r=0$ is 
a singular or a `critical' point. For finite but large $n$, there should 
then be a smooth crossover function interpolating between these two limits. In this
section, these crossover scalings functions are derived analytically
and compared to numerical simulations. 

\subsection{Scaling functions associated with the maximum $m(n)$}

We first consider the crossover scaling functions (near $r\to 0$) associated
with the mean and the variance of the maximum $m(n)$. Let us first focus
on the mean. Our starting point is the exact generating function for the
mean in Eq. (\ref{moment-1}), where we recall that $z_0=z_1(s_{-})$ is
given in Eq. (\ref{z1}). We also remind the reader that
\begin{equation}
a= (1-r)\lambda \quad {\rm and}\quad s_{-}= 
\frac{1}{a}\left(1-\sqrt{1-a^2}\right)\, .
\label{defas_}
\end{equation}
Since, we want to analyze
the behavior of $\langle m(n)\rangle$ for large $n$, and simultaneously
$r\to 0$, we 
need to set
$\lambda$ close to $1$ in Eq. (\ref{moment-1}). We set, $\lambda=1-p$ where $p$ 
is small. Upon inspecting $s_{-}$ in Eq. (\ref{defas_}), it follows that
the right scaling limit is when $p\to 0$, $r\to 0$, keeping the ratio $p/r$ 
fixed. In the real space, this limit corresponds to $r\to 0$, $n\to \infty$
but keeping the product $r\,n$ fixed. 
In this limit, the leading behavior of $s_{-}$ can be easily worked out
to give
\begin{equation}
s_{-} \approx 1 - \sqrt{2(r+p)} \;.
\label{s_leading}
\end{equation}
Substituting this leading behavior of $s_{-}$ in Eq. (\ref{z1}), it follows 
that in the scaling limit
\begin{equation}
z_0= z_1(s_{-})\approx 1 + \frac{\sqrt{2}\,p}{\sqrt{r+p}}\, .
\label{z0scaling}
\end{equation}
We substitute this leading behavior of $z_0$ on the right hand
side (rhs) of Eq. (\ref{moment-1}) and use $\lambda=1-p$.
In the limit $p\to 0$, the sum on the left hand side (lhs) of Eq. 
(\ref{moment-1}) can be approximated by an integral, yielding in
the scaling limit 
\begin{equation}
\int_0^\infty \langle m(n)\rangle\, e^{-p\,n}\, dn \approx \frac{\sqrt{r+p}}{\sqrt{2}\, 
p^2}\, .
\label{meanm_laplace}
\end{equation}
By power counting on both sides of Eq. (\ref{meanm_laplace}), it follows that
$\langle m(n)\rangle$ must have the following scaling behavior
\begin{equation}
\langle m(n)\rangle \approx \sqrt{n}\, f_m(r\,n),
\label{meanm_scaling.2}
\end{equation}
in the appropriate scaling limit $r\to 0$, $n\to \infty$ with the
product $r\,n$ fixed. 

Substituting this scaling behavior on the
lhs of Eq. (\ref{meanm_laplace}), setting $p/r=s$ and making a
change of variable $rn=y$, yields the following equation
for the scaling function $f_m(y)$
\begin{equation}
\int_0^{\infty} \sqrt{y}\, f_m(y)\, e^{-s\,y}\, dy= 
\frac{\sqrt{1+s}}{\sqrt{2}\, s^2}\,.
\label{fmy_laplace}
\end{equation}
To invert the Laplace transform on the rhs of Eq. (\ref{fmy_laplace}), we
reexpress
\begin{equation}
\frac{\sqrt{1+s}}{s^2}= \frac{1}{s^2\sqrt{1+s}}+ \frac{1}{s\sqrt{1+s}}\,.
\label{rational_break}
\end{equation}
Each term on the rhs of Eq. (\ref{rational_break}) is an elementary function
that can be easily inverted using convolution theorem. Inverting, we then 
get an exact expression for the scaling function 
\begin{equation}
f_m(y)= \frac{1}{\sqrt{2y}}\left[\left(y+\frac{1}{2}\right)\, {\rm erf}
\left(\sqrt{y}\right)+ \sqrt{\frac{y}{\pi}}\, e^{-y}\right]\, ,
\label{fmy.1}
\end{equation}
where ${\rm erf}(z)= \frac{2}{\sqrt{\pi}}\, \int_0^z e^{-u^2}\, du$. The 
function $f_m(y)$ has the following asymptotic behaviors
\begin{equation}
f_m(y) \sim 
\begin{cases} \displaystyle
\sqrt{\frac{2}{\pi}} + O(y) &\text{as $y\to 0$, } \\[5mm]
\displaystyle
\sqrt{\frac{y}{2}}+ O\left(\frac{1}{\sqrt{y}}\right) 
& \text{as
$y\to \infty$.}
\end{cases}
\label{fmy_asymp}
\end{equation}

\begin{figure}
\centerline{\includegraphics[width=0.95\hsize]{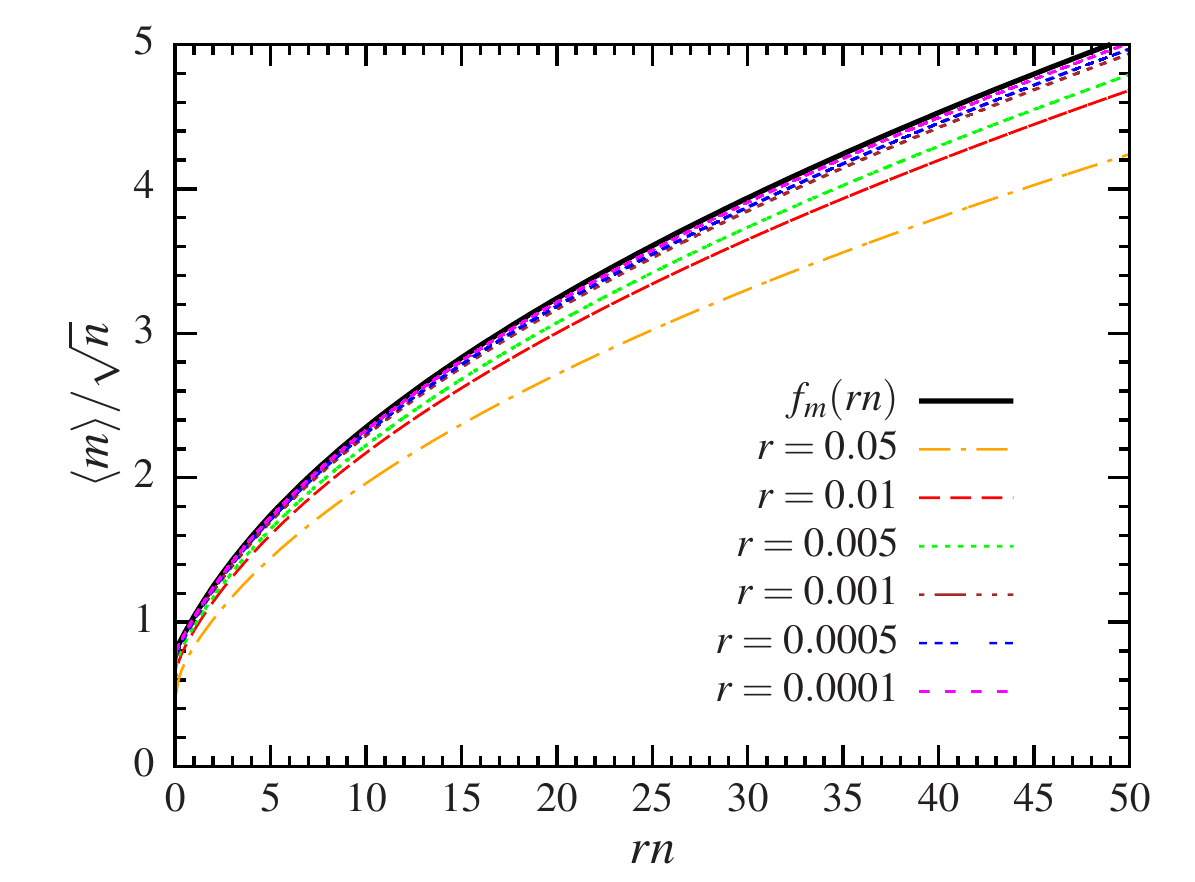}}
\caption{\label{mean-max} (Color online) $\frac{\langle m(n)\rangle}{\sqrt{n}}$
plotted vs. $y=r\,n$ for different small values of $r$. As $r\to 0$, the curves
approach the analytical scaling function $f_m(y)$ in Eq. (\ref{fmy.1}) plotted
as a solid line.} 
\end{figure}

Thus, when $r=0$, using $f_m(0)=\sqrt{2/\pi}$ in Eq.
(\ref{meanm_scaling.2}) yields the asymptotic behavior of the mean,
$\langle m(n)\rangle \simeq \sqrt{2\,n/\pi}$. In contrast, when $r>0$,
as $n\to \infty$, the scaling argument $y=rn\to \infty$. Hence, using
the other asymptotic behavior in Eq. (\ref{fmy_asymp}) as $y\to
\infty$, yields the linear growth $\langle m(n)\rangle \simeq
\sqrt{r/2}\, n$. Note that the speed $v(r)$ in Eq. (\ref{v(r)}) indeed
tends to $v(r)\to \sqrt{r/2}$ as $r\to 0$.  The exact scaling function
$f_m(y)$ thus interpolates smoothly between these two limits. For any
small but nonzero $r$, we thus expect that $\langle m(n)\rangle$, as a
function of $n$, will initially grow as $\sim \sqrt{\frac{2\,n}{\pi}}$
(the critical behaviour at $r=0$), before crossing over at a
characteristic time $n^*(r)\sim 1/r$ to the off-critical linear
growth, $\langle m(n)\rangle \sim \sqrt{r/2}\, n$.  In
\Fref{mean-max}, we compare the numerical simulation results for small
values of $r$ and show how they approach the analytical scaling
function $f_m(y)$ as $r\to 0$.  \\

Next we consider the variance of the maximum, $\sigma_m^2= \langle 
m^2(n)\rangle-
{\langle m(n)\rangle}^2$ in the scaling limit $r\to 0$, $n\to \infty$ but
keeping the product $r\,n$ fixed. For this, we now need to analyze the
second moment in Eq. (\ref{moment-2}) in the scaling limit. The analysis
proceeds more or less as in the case of the mean. We do not repeat this
computation here and just mention the final result. In the scaling
limit $r\to 0$, $n\to \infty$ with the product $rn$ fixed, we find
that the variance behaves as
\begin{equation}
\sigma_m^2\approx  n \, F_m(r\,n),
\label{varm_scaling.2}
\end{equation}
with the scaling function $F_m(y)$ given by
\begin{equation}
F_m(y)= 1+\frac{y}{2}- f_m^2(y)\, ,
\label{Fmy.1}
\end{equation}
where $f_m(y)$ is given in Eq. (\ref{fmy.1}). The scaling function has the 
following asymptotic behaviors
\begin{equation}
F_m(y) \to 
\begin{cases} \displaystyle
1-\frac{2}{\pi} &\text{as $y\to 0$}, \\[3mm]
\displaystyle
\frac{1}{2} &\text{as $y\to \infty$}.
\end{cases}
\label{Fmy_asymp}
\end{equation}

\begin{figure}
\centerline{\includegraphics[width=0.95\hsize]{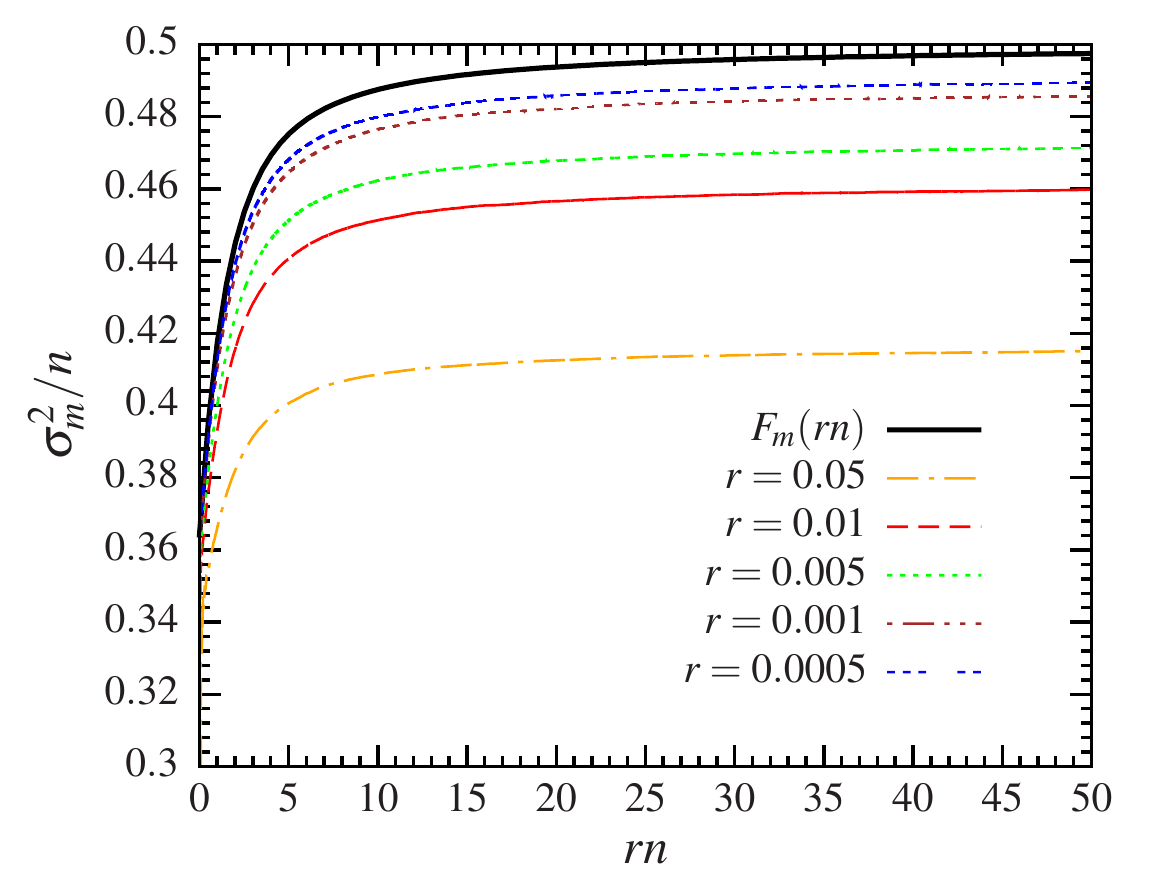}}
\caption{\label{var-max} (Color online) $\sigma_m^2/n$
plotted vs. $y=r\,n$ for different small values of $r$. As $r\to 0$, the curves
approach the analytical scaling function $F_m(y)$ in Eq. (\ref{Fmy.1}) plotted
as a solid line.}
\end{figure}

Hence, for $r=0$, using $F_m(0)= 1-\frac{2}{\pi}$ in Eq. (\ref{varm_scaling.2})
gives the asymptotic behavior of the variance, $\sigma_m^2\simeq 
\left(1-\frac{2}{\pi}\right)\, n$, i.e., the result for normal diffusion 
without
resetting. In contrast, for $r>0$, as $n\to \infty$, the scaling argument
$y\to \infty$. Hence, using $F_m(y)\to 1/2$ as $y\to \infty$ in Eq. 
(\ref{varm_scaling.2}) gives, $\sigma_m^2\simeq n/2$. This agrees
perfectly with the finite $r$ result, $\sigma_m^2\simeq D(r)\,n$
with $D(r)$ given in Eq. (\ref{D(r)}), since $D(r\to 0)=1/2$.
The exact scaling function $F_m(y)$ thus interpolates smoothly
between these two limits. In \Fref{var-max}, we compare the
numerical simulation results with the analytical scaling function $F_m(y)$
in Eq. (\ref{Fmy.1}).
\\

\subsection{Scaling functions associated with the maximum $x(n)$}

We now turn to the scaling behavior of the mean and the variance of the 
position $x(n)$. We start with the mean whose exact generating function
is given in Eq. (\ref{m1x}). To analyze the scaling limit
$r\to 0$, $n\to \infty$ while keeping the product $y=r\,n$ fixed, we follow
the same procedure as in the case of the maximum. Setting $\lambda=1-p$ with
$p\to 0$, and using Eq. (\ref{z0scaling}), we find that Eq. (\ref{m1x}) 
reduces, in the scaling limit, to the following integral
\begin{equation}
\int_0^\infty \langle x(n)\rangle\, e^{-p\,n}\, dn \approx 
\frac{r}{p^2\, \sqrt{2 (r+p)}}\, .
\label{meanx_laplace}
\end{equation}
It then indicates the following scaling behavior for the mean position
\begin{equation}
\langle x(n)\rangle \approx \sqrt{n}\, f_x(r\,n),
\label{meanx_scaling.2}
\end{equation} 
where $f_x(y)$, using Eq. (\ref{meanx_laplace}), satisfies
\begin{equation}
\int_0^{\infty} \sqrt{y}\, f_x(y)\, e^{-s\,y}\, dy=
\frac{1}{s^2\, \sqrt{2(1+s)}}\,.
\label{fxy_laplace}
\end{equation}
One can again easily invert the Laplace transform in \eref{fxy_laplace}
to get
\begin{equation}
f_x(y)  =
\frac{1}{\sqrt{2y}}\left[\left(y-\frac{1}{2}\right)\, {\rm erf}
\left(\sqrt{y}\right)+ \sqrt{\frac{y}{\pi}}\, e^{-y}\right]\, .
\label{fxy.1}
\end{equation}
It has the asymptotics
\begin{equation}
f_x(y)\sim 
\begin{cases}\displaystyle
\frac{2}{3}\sqrt{\frac{2}{\pi}}\, y + O\left(y^2\right) & \text{as
  $y\to 0$,} \\[5mm] \displaystyle
\sqrt{\frac{y}{2}}+O\left(\frac{1}{\sqrt{y}}\right) &\text{as $y\to
  \infty$}.
\end{cases}
\label{fxy_asymp}
\end{equation}

\begin{figure}
\centerline{\includegraphics[width=0.95\hsize]{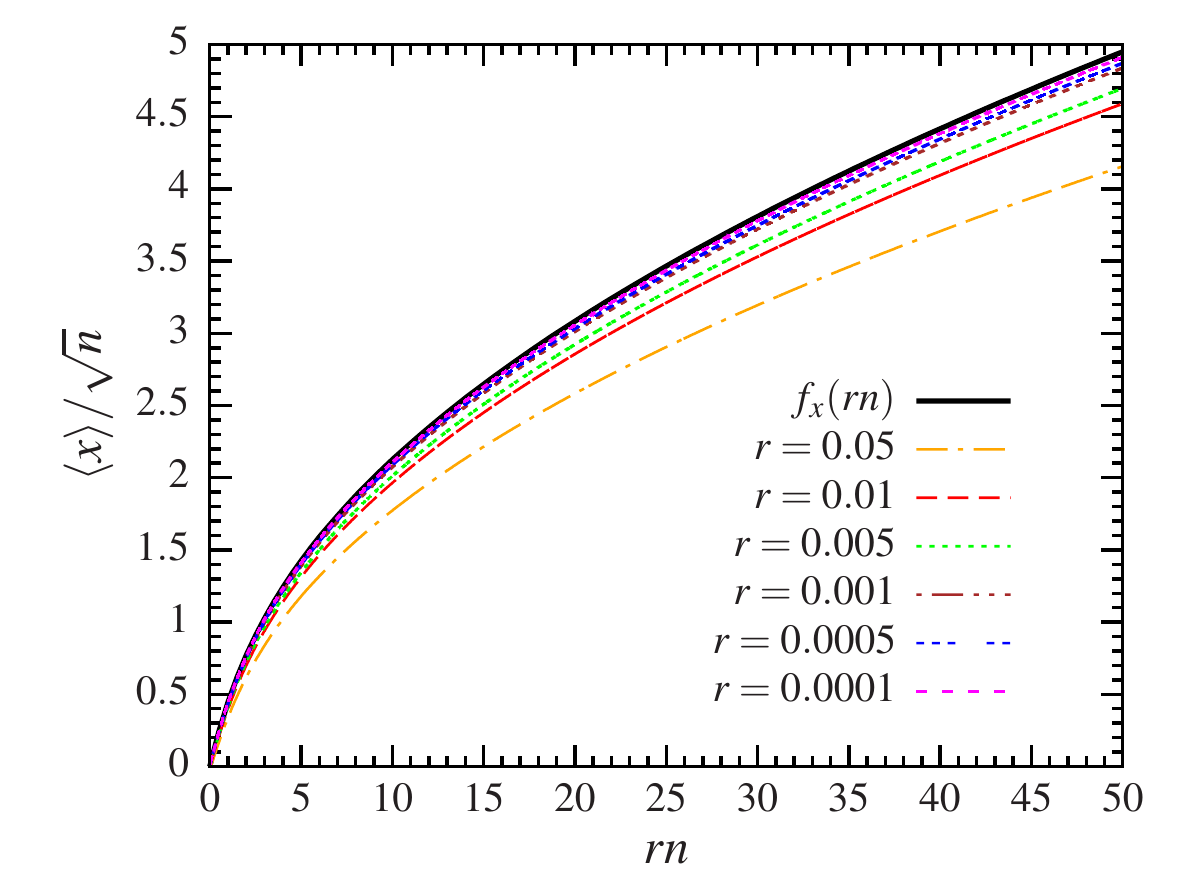}}
\caption{\label{mean-pos} (Color online) $\frac{\langle x(n)\rangle}{\sqrt{n}}$
plotted vs. $y=r\,n$ for different small values of $r$. As $r\to 0$, the curves
approach the analytical scaling function $f_x(y)$ in Eq. (\ref{fxy.1}) plotted
as a solid line.}
\end{figure}

When $r=0$, using $f_x(0)=0$, one recovers 
the standard random walk (without resetting) result, $\langle x(n)\rangle =0$.
In contrast, for $r>0$, when $n\to \infty$, i.e., the product $y=r\,n\to 
\infty$, using the large $y$ asymptotic behavior in Eq. (\ref{fxy_asymp}),
one gets, $\langle x(n)\rangle \simeq \sqrt{r/2}\, n$, compatible with
the linear growth with speed $v(r)$ in Eq. (\ref{rpmeanx}) upon noting
that $v(r\to 0)=\sqrt{r/2}$. The scaling function $f_x(y)$ interpolates between
these two limits. \Fref{mean-pos} demonstrates how simulation results
converge to the analytical scaling function $f_x(y)$ in Eq. (\ref{fxy.1})
as $r\to 0$.
\\

\begin{figure}
\centerline{\includegraphics[width=0.95\hsize]{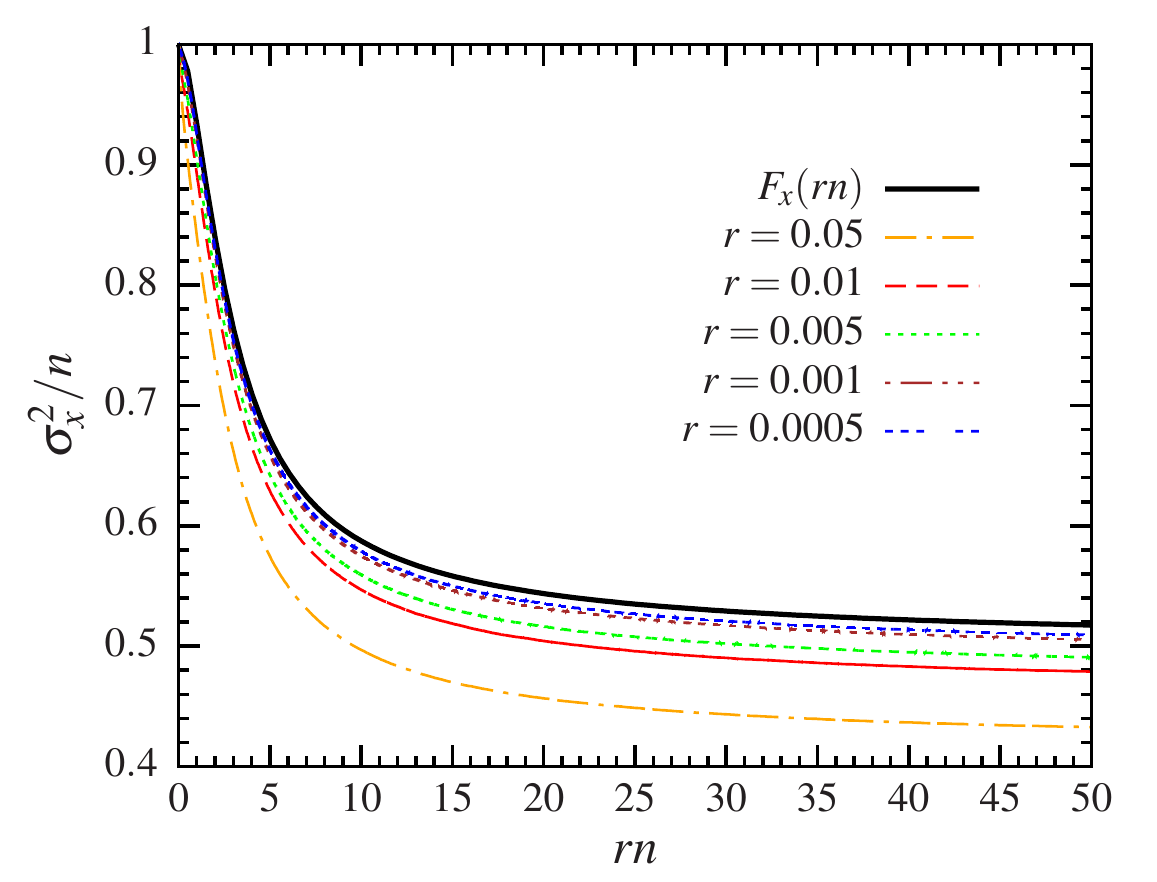}}
\caption{\label{var-pos} (Color online) $\sigma_x^2/n$
plotted vs. $y=r\,n$ for different small values of $r$. As $r\to 0$, the curves
approach the analytical scaling function $F_x(y)$ in Eq. (\ref{Fxy.1}) plotted
as a solid line.}
\end{figure}

We next consider the scaling behavior of the variance of the position,
$\sigma_x^2= \langle x^2(n)\rangle - {\langle x(n)\rangle}^2$, in the
scaling limit $r\to 0$, $n\to \infty$ with the product $y=r\,n$ fixed.
Here we analyze the generating function for the second moment in Eq. 
(\ref{m2x}) in the scaling limit. Since the procedure is identical
as in the case of the maximum, we skip the details and present only the
result. We find that in the scaling limit, the variance of the position behaves
as
\begin{equation}
\sigma_x^2\approx  n \, F_x(r\,n),
\label{varx_scaling.2}
\end{equation}
with the scaling function $F_x(y)$ given by
\begin{equation}
F_x(y)= \frac{y}{2}+\frac{1-e^{-y}}{y}- f_x^2(y)\, ,
\label{Fxy.1}
\end{equation}
where $f_x(y)$ is given in Eq. (\ref{fxy.1}). The scaling function $F_x(y)$ has 
the
following asymptotic behaviors
\begin{equation}
F_x(y) \to 
\begin{cases} 
1  &\text{as  $y\to 0$,} \\
\frac{1}{2}  &\text{as $y\to \infty$.}
\end{cases}
\label{Fxy_asymp}
\end{equation}
Using these asymptotic behaviors, it is again easy to check
that $F_x(y)$ 
smoothly interpolates between the critical $(r=0)$ and the off-critical
$(r>0)$ behavior of the variance of the position, for finite but large $n$. 
In \Fref{var-pos}, we compare the
numerical simulation results with the analytical scaling function $F_x(y)$
in Eq. (\ref{Fxy.1}).
\\

\section{Conclusion}
In conclusion, we have considered a model of random walk in one 
dimension where the walker, at
each time step, resets to the maximum of the already visited positions
with a certain probability $r$. For $r=0$, it reduces to the standard
random walk in one dimension. The presence of a nonzero resetting
probability $r$ changes drastically the asymptotic behavior of
the walker. 
We find that on average, both the
position and the maximum move with the same speed, and we have
obtained an exact expression for this speed $v(r)$. The fluctuations
about the mean is again described by the same diffusion coefficient
$D(r)$ for both. We also obtain the large deviation form of the
probabilities of finding the walker at a distance $y$ away from the
maximum. The associated large deviation function shows a second order
phase transition.

An interesting extension would be to study the walk which resets
either to the maximum or the minimum with an equal probability
$r/2$. In general, one could ask the question in higher dimension,
where the walker resets to one of the boundary sites of the already
visited sites. Yet another extension, beyond the study of the fluctuations of
the position and the maximum studied here, concerns the study of the search process with resetting to the maximum. 
For instance, it would be interesting
to compute the mean first-passage time to an immobile or a moving target 
for a such a random walker submitted to random resetting to the maximum.   

\label{conclusion}

\begin{acknowledgements}
The authors acknowledge the support of the Indo-French Centre for the
Promotion of Advanced Research (IFCPAR/CEFIPRA) under Project 4604-3. 
S.~N.~M acknowledges useful discussions with P. Sen and her hospitality at
the physics department of Calcutta University, 
where part of this work was done during a visit in July, 2015.
\end{acknowledgements}


\end{document}